\documentclass[a4paper,twocolumn,english]{revtex4-1}
\usepackage[T2A,T1]{fontenc}
\usepackage[latin9]{inputenc}
\usepackage{color}
\usepackage{array}
\usepackage{rotating}
\usepackage{float}
\usepackage{multirow}
\usepackage{amsmath}
\usepackage{amssymb}
\usepackage{graphicx}

\makeatletter

\pdfpageheight\paperheight
\pdfpagewidth\paperwidth

\newcommand{\noun}[1]{\textsc{#1}}
\DeclareRobustCommand{\cyrtext}{%
  \fontencoding{T2A}\selectfont\def\encodingdefault{T2A}}
\DeclareRobustCommand{\textcyr}[1]{\leavevmode{\cyrtext #1}}
\AtBeginDocument{\DeclareFontEncoding{T2A}{}{}}

\newcommand{\lyxmathsym}[1]{\ifmmode\begingroup\def\b@ld{bold}
  \text{\ifx\math@version\b@ld\bfseries\fi#1}\endgroup\else#1\fi}

\providecommand{\tabularnewline}{\\}

\makeatother

\usepackage{babel}
\begin{document}

\title{Self-consistent mapping: Effect of local environment on formation
of magnetic moment in $\boldsymbol{\alpha\mathit{-FeSi_{2}}}$.}

\author{V.S. Zhandun}
\email{jvc@iph.krasn.ru}
\thanks{Corresponding author}
\affiliation{Kirensky Institute of Physics, Federal Research Center \textquotedbl{}Krasnoyarsk
Science Centre, Siberian Branch of the Russian Academy of Sciences'',
660036 Krasnoyarsk, Russia}
\author{N.G. Zamkova, S.G. Ovchinnikov, and I.S. Sandalov}

\affiliation{Kirensky Institute of Physics, Federal Research Center \textquotedbl{}Krasnoyarsk
Science Centre, Siberian Branch of the Russian Academy of Sciences'',
660036 Krasnoyarsk, Russia}

\begin{abstract}
The Hohenberg-Kohn theorem establishes a basis for mapping of the
exact energy functional to a model one provided that their charge
densities coincide. We suggest here to use a mapping in a similar
spirit: the parameters of the formulated multiorbital model should
be determined from the \textcolor{black}{requirement that the self-consistent
charge and spin densities found from the }\textcolor{black}{\emph{ab
initio}}\textcolor{black}{{} and model calculations have to be as close
to each other as possible. The analysis of the model allows for detailed
understanding of the role played by different parameters of the model
in the physics of interest. After finding the areas of interest in
the phase diagram of the model we return to the}\textcolor{black}{\emph{
ab initio}}\textcolor{black}{{} calculations and check if the effects
discovered are confirmed or not. Because of the last controlling step
we call this approach as hybrid self-consistent mapping approach (HSCMA).
As an example of the approach we present the study of }the effect
of silicon atoms substitution by the iron atoms and \emph{vice versa}
on the magnetic properties in the iron silicide $\alpha-FeSi_{2}$.
The DFT+GGA calculations are mapped to the model\textcolor{black}{{}
with intraatomic Coulomb and exchange interactions, hoppings to nearest
and next nearest atoms and exchange of the delocalized electrons between
iron atoms; the magnetic moments on atoms and charge densities of
the material are found self-consistently within the Hartree-Fock approximation.}

We find that while the stoichiometric $\alpha-FeSi_{2}$ is nonmagnetic,
the substitutions generate different magnetic structures. For example,
the substitution of three $Si$ atoms by the $Fe$ atoms results in
the fe\textcolor{black}{rrima}gnetic structure whereas the substitution
of four $Si$ atoms by $Fe$ atoms gives rise to either the nonmagnetic
or the ferromag\textcolor{black}{netic state depending on the type
of local enviroment of the substitutional }$Fe$ \textcolor{black}{atoms.
Besides, contrary to the commonly accepted statement that the destruction
of the magnetic moment is controlled only by the number of $Fe-Si$
nearest neighbors, we find that actually it} is controlled by the
$Fe-Fe$ \emph{next}-nearest-neighbors' hopping parameter.\textcolor{black}{{}
This finding led us to the counterintuitive conclusion: }an \emph{increase
of Si concentration} in $Fe_{1-\boldsymbol{x}}Si_{2+\boldsymbol{x}}$
ordered alloys may lead to a ferromagnetism\textcolor{black}{. This}
conclusion is confirmed by the calculation within GGA-to-DFT.
\end{abstract}


\maketitle

\section{I\noun{NTRODUCTION}}

The method of mapping of first-principle density functional theory
(DFT) calculations to the effective Heisenberg model for theoretical
study of the magnetic properties of solids was developed in the series
of works \cite{key-34}. The role played by the electronic subsystem
in this approach is reduced to formation of the lowest-order pairwise
effective exchange interaction of classical spins. In order to have
an opportunity to use the well-developed many-body perturbation theory
and to obtain the physical picture of the formation of the magnetic
and, especially, the non-magnetic properties of the matter by the
electronic subsystem one either have to use Hedin's GW approximation\cite{key-35}
or a detailed model which includes all atoms, their key orbitals and
symmetry of the lattice in question, hopping parameters and Coulomb
interactions. The GW approximation (even without the vertex corrections)
is extremely time and computer-resources consuming. For this reason
the route with more simple model Hamiltonians seems to be more efficient
for highlighting the physics.

The construction of the hopping parameters for certain symmetries
has been described by Slater and Koster\cite{key-23}. Then, the phase
diagram for the chosen model in the multidimensional space of these
hopping parameters, Coulomb and exchange matrix elements can be constructed
in a proper approximation. However, one point in this multi-parameter
space corresponds to each real material, which can be described by
such a model. A change of the external conditions for the material,
like applying a pressure, temperature, or placing a film of the materaial
on some substrate, will move this point from the initial position
only slightly. Therefore, in order to be able to predict the behavior
of real material, we have to know the material-in-question coordinates
in the parameter space with good accuracy. Unfortunately, a unique
receipt how to find the position of the material in the model parameter
space does not exist. Here we suggest the following way to resolve
this difficulty. Since the DFT-based calculations usually give a reasonably
good description of metals and produce corresponding self-consistent
spin- and charge densities, we can speculate in the DFT spirit: we
use the requirement that the system with a model Hamiltonian has the
spin- and charge densities as close as possible (ideally, the same)
to the one obtained within the first-principle calculations for finding
the parameters of the model. Then, having obtained some prediction
within the model calculations we return to the first-principle ones
in order to check the validity of the model prediction. This is an
essence of the suggested here\textcolor{black}{{} }\textcolor{black}{\emph{hybrid
self-consistent mapping approach}}\textcolor{black}{{} (HSCMA). }It
may seem that such approach should work only within the validity domain
of the chosen approximation for the DFT (in our case GGA-to-DFT).
However, the largest contribution to the energy of the system and
formation of the local charge density $\rho_{DFT}(r)$ comes from
the Hartree part of interaction, which is treated in DFT pretty well.
That is why we want to start \emph{at least} from the point in the
model parameter space which provides $\rho_{model}(r)$ close to $\rho_{DFT}(r)$.
Further the model with these parameters can be used for description
of the phenomena beyond reach of DFT. Then the question arises why
we want that the \emph{self-consisten}t model and DFT charge densities
(magnetic moments) have to be close to each other? The matter is that
we want to know the \emph{bare} parameters of the model in order to
be able to use the diagrammatic methods for dressing them and to avoid
double counting. The most difficult question here is to find an approximation
for the model calculation which would correspond to the one used in
DFT. It clear, however, that the constraints in the accuracy of both
methods allow to require a rough correspondence of charge densities.
Indeed, on the one hand, in the DFT we use the exchange-correlation
potential with restricted and, often, unknown validity domain (moreover,
it is known that the calculations within the same approximation, but
different packages, produce non-coinciding results). On the other
hand, the model has to contain much smaller number of the interactions
and hopping parameters (otherwise we will come to GW type of description
at least). In spite of this uncertainty at the initial step this approach
is attractive because it does not contain any fitting parameters,
which should be taken from experiment. From this point of view it
should be considered as first-principle approach. We have chosen to
treat the model within the Hartree-Fock approximation.

The HSCMA is applied here for the analysis of the magnetic properties
of $\alpha-FeSi_{2}$ -based ordered alloys.

The growth of \emph{Fe}-silicides on silicon has been widely studied
in recent years because, depending on their phase, crystal structure
and composition, they can be semiconducting, metallic and/or ferromagnetic,
and hence offer a large variety of potential applications when integrated
into silicon-based devices \cite{key-1}-\cite{key-3}. To this day,
several \emph{Fe}\textendash silicide structures have been reported.
At the \emph{Fe}-rich side of the binary phase diagram, metallic as
well as ferromagnetic $Fe_{5}Si_{3}$ and $Fe_{3}Si$ ($DO_{3}$ structure)
\cite{key-4,key-5}have already been established as key materials
for spintronics \cite{key-5,key-6}. The \emph{Si}-rich side of the
phase diagram contains several variants of a disilicide stoichiometric
compound, such as the high-temperature tetragonal metallic \emph{$\alpha-FeSi_{2}$}
phase \cite{key-8}, with applications as an electrode or an interconnect
material \cite{key-9,key-10}, and the orthorhombic semiconducting
\emph{$\beta-FeSi_{2}$} phase \cite{key-11}, which due to its direct
band gap is an interesting candidate for thermoelectric, photovoltaic
and optoelectronic devices \cite{key-12}. While room-temperature
stable $\beta$-phase is well-studied, tetragonal $\alpha$-phase
do not attract great interest until recently. This is due to this
phase is metastable and exist only at temperatures above $950^{\mathrm{o}}C$
\cite{key-8}. However, the iron silicides, which do not exist in
bulk, can be stabilized as films. In Refs. \cite{key-9},\cite{key-13}-\cite{key-16}
a successful fabrication of thin films \emph{$\alpha-FeSi_{2}$} was
reported. Also, while the magnetic order is not observed in bulk stoichiometric
disilicide \emph{$\alpha-FeSi_{2}$}, ferromagnetism was found \cite{key-16}
in the metastable phase \emph{$\alpha-FeSi_{2}$}, which was stabilized
in epitaxial-film grown on the silicon substrate. The authors of Refs.
\cite{key-10}, \cite{key-17} reported that the magnetic moments
on Fe atoms $\mu=1.8\mu_{B}$ \cite{key-10} and $\mu=3.3\mu_{B}$
\cite{key-17} in \emph{$\alpha-FeSi_{2}$} nanoislands and nano-stripes
on \emph{Si} (111) substrate arise. These experimental achievements
have good perspective for the integration of the \emph{$FeSi$} -based
magnetic devices into silicon technology, and, therefore, demands
for the detailed understanding of the physics of the magnetic moment
formation in these compounds.

Traditionally, the appearance of the magnetic structure in \emph{$Fe\lyxmathsym{\textendash}Si$}
alloys is related to the increase of the concentration of \emph{$Fe$}
atoms. So, the unusual ferromagnetism in epitaxial-film form \emph{$\alpha-FeSi_{2}$}
authors \cite{key-16} explain by the appearance of substitutional
\emph{Fe} atoms on \emph{Si} sublattice. According to the \emph{ab
initio} calculation in the framework of \emph{Coherent Potential Approximation}
(CPA) performed in \cite{key-16} the ferromagnetism in thin films
\emph{$\alpha-FeSi_{2}$} appears with the substitution of small percent
of silicon atoms by the iron atoms. Particularly, when the concentration
of substitution \emph{$Fe$} atoms reaches $3.3\%$, these \emph{$Fe$}
atoms acquire magnetic moment $\mu=2.4\mu_{B}$. Similar explanation
of anomalously high total magnetic moment was suggested also by the
authors of Refs. \cite{key-10}, \cite{key-17}. The decrease of the
magnetic moments of \emph{Fe} atoms with the increase of the \emph{Si}
concentration was observed experimentally in the iron silicides $Fe_{x}Si_{1+x}$
and discussed in the framework of the phenomenological local environment
models \cite{key-18}-\cite{key-20}. It was noticed, that the changing
of magnetic moment of \emph{Fe} atom in iron silicides $Fe_{x}Si_{1+x}$,\textcolor{black}{{}
rather depends on the number of }\textcolor{black}{\emph{Si}}\textcolor{black}{{}
atoms in the nearest local environment of iron and not on the concentration
of }\textcolor{black}{\emph{Si}}\textcolor{black}{{} atoms. In} our
work\textcolor{black}{{} \cite{key-21}} the mechanism of magnetic moment
formation in $Fe_{3}Si$ is analysed in the framework of the multiorbital
model, where it is shown that the neighboring \emph{Fe} atoms along
crystallographic axes as well as \emph{Si} atoms in the first coordination
sphere play the crucial role in the destruction of the \emph{Fe} magnetic
moments. Namely, the increase of the number of such \emph{$Fe$} neighbors
leads to the decrease of the \emph{Fe} magnetic moment. Iron atoms
in \emph{$\alpha-FeSi_{2}$} have only silicon atoms as the nearest
neighbors and from the traditional point of view \cite{key-18}-\cite{key-20}
it is naturally to assume that the absence of the magnetism in this
silicide is caused by the nearest silicon environment. However, the
specific feature of the \emph{$\alpha-FeSi_{2}$} structure is the
presence of the alternating \emph{$Fe$} and \emph{$Si$} planes,
which are perpendicular to the tetragonal axis of the cell (Fig.\ref{fig:1a}a).
In such plane \emph{Fe} atoms are surrounded only by \emph{$Fe$}
atoms arranged along the crystallographic axes. Our analysis\textcolor{black}{{}
\cite{key-21} prompts that} such mutual arrangement of \emph{$Fe$}
atoms should results in the magnetic moment destruction. The target
of this work is to investigate the influence of local environment
on the formation of the magnetic moments on iron atoms in the silicide
\emph{$\alpha-FeSi_{2}$,}\textcolor{black}{{} its ordered }\textcolor{black}{\emph{$Fe$}}\textcolor{black}{{}
- rich solid solutions with substitutional }\textcolor{black}{\emph{$Fe$}}\textcolor{black}{{}
atoms $Fe_{1+x}Si_{2-x}$ and }\textcolor{black}{\emph{Si}}\textcolor{black}{{}
- rich one with substitutional }\textcolor{black}{\emph{Si}}\textcolor{black}{{}
atoms, $Fe_{1-x}Si_{2+x}$.} Particularly, we will address the question
about the role, played by second neighbors of \emph{Fe} ions in the
physics of magnetic moment formation.

\textcolor{black}{The paper is organized as follow. In Sec.II we provide
the details of }\textcolor{black}{\emph{ab initio}}\textcolor{black}{{}
and model calculations. The results of the }\textcolor{black}{\emph{ab
initio}}\textcolor{black}{{} calculations of }\textcolor{black}{\emph{$\alpha-FeSi_{2}$}}\textcolor{black}{{}
and its }\textcolor{black}{\emph{Fe}}\textcolor{black}{-rich alloys
are given in Sec. IIIA. The results of the model calculations of }\textcolor{black}{\emph{$\alpha-FeSi_{2}$}}\textcolor{black}{{}
and its }\textcolor{black}{\emph{Fe}}\textcolor{black}{-rich alloys
and the dependence of magnetic moments on the hopping matrix elements
are presented in Sec. IIIB. The results of the}\textcolor{black}{\emph{
ab initio}}\textcolor{black}{{} investigation of }\textcolor{black}{\emph{Si}}\textcolor{black}{-rich
alloys of }\textcolor{black}{\emph{$\alpha-FeSi_{2}$}}\textcolor{black}{{}
are described in Sec. IIIC. Sec.IV contains the summary of the obtained
results and conclusions.}

\section{HSCMA: THE HYBRID \emph{ab initio} AND MODEL CALCULATION METHOD }

In this work we combine the \emph{ab initio} calculations with the
model one. We use the following scheme. First we perform the calculation
of electronic and magnetic properties of the compound of interest
within the framework of \emph{DFT-GGA} for different way of silicon
atoms substitution by iron atoms taking into account the relaxation
of atomic positions. Then we perform mapping the \emph{DFT-GGA} results
to\textcolor{black}{{} the multiorbital model, suggested in Ref.\cite{key-21}.}
The guiding argument for the formulation of the model are: the model
should 1) contain as little as possible parameters; 2) contain the
specific information about the compound in question, \emph{i.e.},
contain proper number of orbitals and electrons, and to posess the
symmetry of the corresponding crystal structure, and 3) contain main
interactions, reflecting our understanding of the underlying physics.
At last, we perfom the mapping following the\emph{ DFT} ideology:
we find the parameters of the model from fitting the its self-consistent
charge density to the one, obtained in \emph{the ab initio} calculations.
The latter step distinguishes our approach from other ones \cite{key-34,key-36,key-37}.
\textcolor{black}{Here we briefly outline the model Hamiltonian, the
details of model calculation are described in \cite{key-21}}. We
include into the Hamiltonian of our model set of interactions between
the \emph{d}-electrons of Fe (\emph{5d}-orbitals per spin) following
Kanamori \cite{key-22}. The structure contains neighboring \emph{$Fe$}
ions, for this reason the interatomic direct \emph{d-d}-exchange and
\emph{d-d}-hopping are included too. The \emph{$Si$} \emph{p}-electrons
(\emph{3p}-orbitals per spin) are modeled by atomic levels and interatomic
hoppings. Both subsystems are connected by \emph{$d-p$}-hoppings.
Thus, the Hamiltonian of the model is:

\begin{equation}
H=H^{Fe}+H_{J'}^{Fe-Fe}+H_{0}^{Si}+H_{hop},
\end{equation}

where

\[
H^{Fe}=H_{0}^{Fe}+H_{K}^{Fe}
\]

\begin{center}
$H_{0}^{Fe}=\sum\varepsilon_{0}^{Fe}\hat{n}_{nm\sigma}^{d};$
\par\end{center}

\begin{center}
$H_{0}^{Si}=\sum\varepsilon_{0}^{Si}\hat{n}_{nm\sigma}^{p};$
\par\end{center}

and the Kanamori\textquoteright s part of the Hamiltonian

\begin{widetext}

\begin{multline}
H_{K}^{Fe}=\frac{U}{2}\sum\hat{n}_{nm\sigma}^{d}\hat{n}_{nm\bar{\sigma}}^{d}+\left(U'-\frac{1}{2}J\right)\sum\hat{n}_{nm}^{d}\hat{n}_{nm'}^{d}\left(1-\delta_{mm'}\right)-\frac{1}{2}J\sum\hat{\boldsymbol{s}}_{nm}^{d}\hat{\boldsymbol{s}}_{nm'}^{d};\\
H_{J'}^{Fe-Fe}=-\frac{1}{2}J'\sum\hat{\boldsymbol{s}}_{nm}^{d}\hat{\boldsymbol{s}}_{n'm'}^{d};\\
H_{hop}=\sum T_{n,n'}^{mm'}p_{nm\sigma}^{\dagger}p_{n'm'\sigma}+\sum t_{n,n'}^{mm'}d_{nm\sigma}^{\dagger}d_{n'm'\sigma}+\sum\left[\left(t'\right)_{n,n'}^{mm'}d_{nm\sigma}^{\dagger}p_{n'm'\sigma}+H.c.\right];\\
\hat{n}_{nm\sigma}^{d}\equiv d_{nm\sigma}^{\dagger}d_{nm\sigma};\;\hat{n}_{nm}^{d}=\hat{n}_{nm\uparrow}^{d}+\hat{n}_{nm\downarrow}^{d};\;\hat{\boldsymbol{s}}_{nm}^{d}\equiv\boldsymbol{\sigma}_{\alpha\gamma}d_{nm\alpha}^{\dagger}d_{nm\gamma};\;\hat{n}_{nm\sigma}^{p}\equiv p_{nm\sigma}^{\dagger}p_{nm\sigma}.\label{eq:H}
\end{multline}

\end{widetext}

Here $p_{nm\sigma}^{\dagger}(p_{nm\sigma})$ and $d_{nm\sigma}^{\dagger}\left(d_{nm\sigma}\right)$
are the creation (annihilation) operators of \emph{$p$}-electrons
on \emph{Si-} and \emph{d}-electrons on \emph{Fe} -ions; $n$ is complex
lattice index, (site, basis); $m$ labels the orbitals; $\sigma$
is spin projection index; $\boldsymbol{\sigma}_{\alpha\gamma}$ are
the Pauli matrices; $U,\:U'=U-2J$ and $J$ are the intraatomic Kanamori
parameters; $J'$ is the parameter of the intersite exchange between
nearest \emph{Fe} atoms. At last, $T_{n,n'}^{mm'},\;t_{n,n'}^{mm'}\;\left(t'\right)_{n,n'}^{mm'}$
are hopping integrals between \emph{$Si-Si$, $Fe-Fe$} and \emph{$Fe-Si$}
atoms, correspondingly. The dependences of hopping integrals $T_{n,n'}^{mm'},\;t_{n,n'}^{mm'}\;\left(t'\right)_{n,n'}^{mm'}$
\textcolor{black}{of }$\boldsymbol{\mathbf{k}}$ were obtained from
the Slater and Koster atomic orbital scheme \cite{key-23} in the
two-center approximation using basic set consisting of five \emph{3d}
orbitals for each spin on each \emph{Fe} and three \emph{3p} orbital
for each spin on each \emph{Si}. In this two-centre approximation
the hopping integrals depend on the distance $\mathbf{R=(\mathit{l}x+\mathit{m}y+\mathit{n}z)}$
between the two atoms, where \textbf{\textcolor{black}{$\mathbf{x,\:}\mathbf{y,}\:\mathbf{z}$}}
are the unit vectors along cubic axis and \emph{l, m, n} are direction
cosines. Then, within the two-center approximation, the hopping integrals
are expressed in terms of Slater \textendash{} Koster parameters $t_{\sigma}=(dd\sigma)$,
$t_{\pi}=(dd\pi)$ and $t_{\delta}=(dd\delta)$ for \emph{$Fe\lyxmathsym{\textendash}Fe$}
hopping, $t_{\sigma}=(pd\sigma)$, $t_{\pi}=(pd\pi)$ for \emph{$Fe-Si$}
and $t_{\sigma}=(pp\sigma)$, $t_{\pi}=(pp\pi)$ for \emph{$Si\lyxmathsym{\textendash}Si$}
hoppings ($\sigma,\:\pi,\:\delta$ specifies the components of the
angular momentum relative to the direction $\mathbf{R}$). Their $\mathbf{k}$-dependence
are given by the functions $\gamma_{\sigma}(\mathbf{k}),\:\gamma_{\pi}(\mathbf{k})$
and $\gamma_{\delta}(\mathbf{k})$, where $\gamma(\boldsymbol{\mathbf{k}})=\sum_{\mathbf{R}}e^{i\mathbf{kR}}$.
The expressions for hopping integrals can be obtained in Table I from
\cite{key-23}. For example, $t_{Fe-Fe}^{xy,xy}(\mathbf{k})=2t_{\pi}\left[cos\left(R_{x}k_{x}\right)+cos\left(R_{y}k_{y}\right)\right]+2t_{\delta}cos\left(R_{z}k_{z}\right)$,
etc. The number of points in the Brillouin zone was taken 1000. Monkhorst-Pack
scheme \cite{key-24} was used for generation of the k-mesh. The model
is solved within the Hartree-Fock approximation (HFA). The band structure
arises due to hopping parameters, which connect nearest neighbors
(NN) and next NN (NNN) sites. The calculations were performed for
three initial states: ferromagnetic (FM), antiferromagnetic (AFM)
and paramagnetic (PM) states. After achieving self-consistency the
state with minimal total energy was chosen. The last step was done
with the help of the Galitsky-Migdal formula for total energy ((10)
in \cite{key-21}), which we adopted for our model.

All \emph{ab initio} calculations presented in this paper have been
performed using the \emph{Vienna ab initio simulation package} (VASP)
\cite{key-25} with projector augmented wave (PAW) pseudopotentials
\cite{key-26}. The valence electron configurations $3d^{6}4s^{2}$
are taken for \emph{$Fe$} atoms and $3s^{2}3p^{2}$ for \emph{Si}
atoms. The calculations is based on the density functional theory
where the exchange-correlation functional is chosen within the Perdew-Burke-Ernzerhoff
(PBE) parametrization \cite{key-27} and the generalized gradient
approximation (GGA) has been used. Throughout all calculations, the
plane-wave cutoff energy is 500eV, and Gauss broadening with smearing
0.05eV is used. The Brillouin-zone integration is performed on the
grid Monkhorst-Pack \cite{key-24} special points $8\times8\times6$.
The optimized lattice parameters\textcolor{black}{{} and atom's coordinates
}were obtained by minimizing the full energy.

\section{RESULTS AND DISCUSSION}

\subsection{\emph{Ab initio} calculations}

\begin{figure}
\begin{tabular}{c}
\multicolumn{1}{c}{\includegraphics[bb=0bp 0bp 640bp 310bp,scale=0.37]{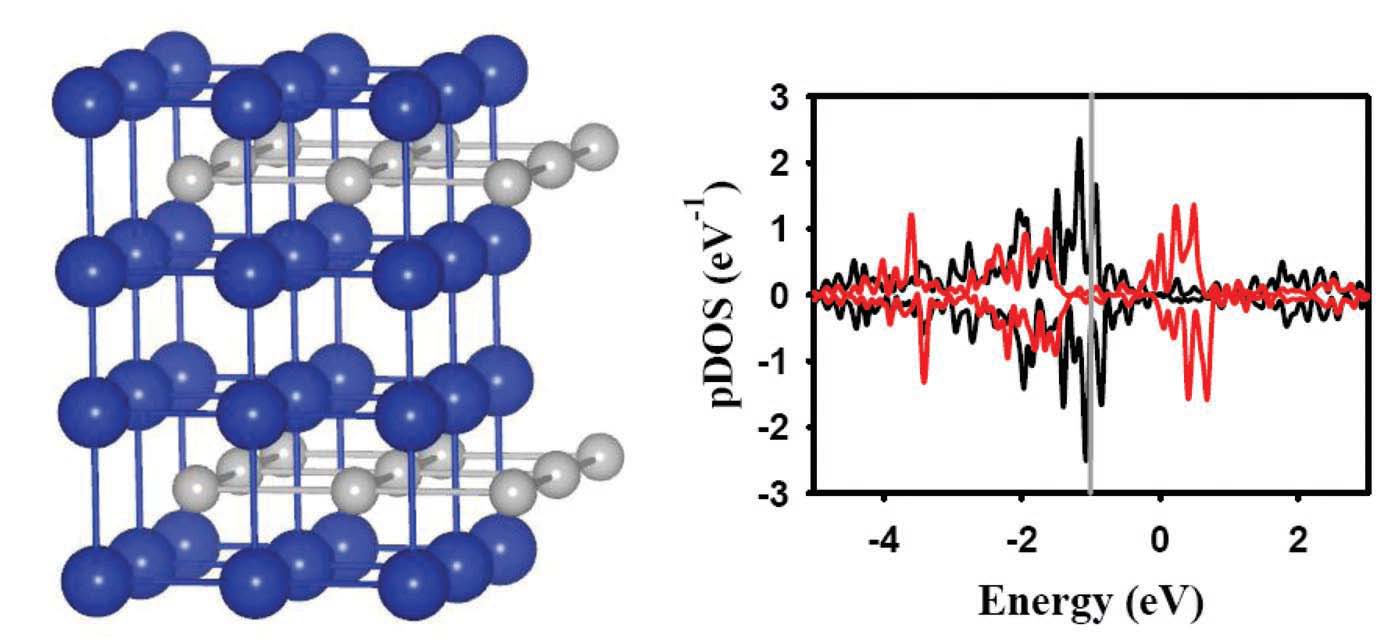}}\tabularnewline
\tabularnewline
\end{tabular}

\caption{\textcolor{black}{\emph{(Color online)}}\textbf{\textcolor{black}{\emph{
}}}Left panel: the structure of \emph{$\alpha-FeSi_{2}$}; \emph{$Si$}
atoms are shown by blue balls, \emph{$Fe$} atoms \textendash{} by
grey balls. Right panel: partial density of electronic states (pDOS)
of \emph{$Fe$} atoms; black line shows $t_{2g}$ states, red line
shows $e_{g}$ states. Zero on the energy axis is chosen at the Fermi
energy. \label{fig:1a}}
\end{figure}

Stoichiometric compound \emph{$\alpha-FeSi_{2}$} has tetragonal space
symmetry group $P4/mmm$ with one formula unit per cell. \textcolor{black}{The
structure is shown in Fig.\ref{fig:1a}}\textcolor{black}{\emph{a}}\textcolor{black}{.
The compound is nonmagnetic metal with lattice parameters from our
}\textcolor{black}{\emph{ab}}\textcolor{black}{{} }\textcolor{black}{\emph{initio}}\textcolor{black}{{}
calculations $a=2.70$Å, $c=5.13$Å that are in a good agreement with
experimental values \cite{key-28}. The structure of }\textcolor{black}{\emph{$\alpha-FeSi_{2}$}}\textcolor{black}{{}
consists of alternating planes of iron and silicon atoms }\textcolor{black}{\emph{$Fe-Si-Si-Fe$}}\textcolor{black}{,
which are perpendicular to the tetrago}nal axis of the cell. Iron
atoms are surrounded by 8 silico\textcolor{black}{n atoms ($R_{Fe-Si}=2.36$Å)
located in the corners of slightly distorted in {[}001{]} direction
cube, the next nearest neighbors (NNN) of iron atoms are }\textcolor{black}{\emph{Fe}}\textcolor{black}{{}
atoms arranged along crystallographic axes }\textbf{\textcolor{black}{x}}\textcolor{black}{{}
and }\textbf{\textcolor{black}{y,}}\textcolor{black}{{} forming the}
iron plane ($R_{Fe-Fe}=2.70$Å). The full density of states of \emph{$\alpha-FeSi_{2}$}
was calculated in the works \cite{key-29}-\cite{key-31} and in our
recent work \cite{key-32}, thus, in the present paper we give only
partial spin-projected density (pDOS) of \emph{Fe d -} electron states
in Fig.\ref{fig:1a}b. As seen, both $t_{2g}$ and $e_{g}$ electrons
are delocalized in a wide energy range and a magnetism is absent.

\textcolor{black}{However, as was mentioned in Introduction, several
recent studies \cite{key-10,key-16,key-17} discovered that a ferromagnetic
state arises in the films of }\textcolor{black}{\emph{$\alpha-FeSi_{2}$}}\textcolor{black}{.
The explanation of the emergence of the magnetic structure, suggested
in these works, is within the commonly accepted opinion, that the
magnetism arises due to an increase of }\textcolor{black}{\emph{Fe}}\textcolor{black}{{}
concentration in the material. The used theoretical approaches, CPA
in the Ref. \cite{key-16} and phenomenological local environment
models in Refs \cite{key-10,key-17}, take into account, however,
only a part of the local environment effects because full account
of them is beyond the reach of the standard CPA methods by construction,
whereas the local environment models \cite{key-18,key-19} take into
account the nearest enviroment only. In Ref.\cite{key-21} we found
that the next nearest environment (NNN) plays a crucial role in the
magnetic moment formation. This motivates us to include the NNN local
environment effects into study of the magnetic properties of }\textcolor{black}{\emph{$Fe$}}\textcolor{black}{{}
- rich ordered alloys both in the framework of DFT calculations and
subsequent analysis in the suggested multiorbital model too. The different
local environment of iron atoms was set by the different spatial arrangement
and number of substitutional }\textcolor{black}{\emph{Fe}}\textcolor{black}{{}
atoms in the ordered alloys $Fe_{1+x}Si_{2-x}$. I}n this part of
paper we presented the results of our \emph{ab initio} calculations
of some of ordered alloys\textcolor{black}{{} $Fe_{1+x}Si_{2-x}$}.
We used for the calculations the supercell $2a\times2a\times c$,
where \emph{a} and \emph{c} \textendash{} the lattice parameters of
stoichiometric \emph{$\alpha-FeSi_{2}$}.

The ordered alloys considered in the present work are shown in Table
\ref{tab:Table 1}. Alloys \textbf{A} and \textbf{B} contain one and
three substitutional \emph{$Fe$} atoms at the $Si$ sites in the
\emph{$Si$}-planes, correspondingly. In the last three alloys \textbf{C},
\textbf{D}, \textbf{E} four \emph{$Si$} atoms were replaced by \emph{$Fe$}
atoms in different ways: in the plane perpendicular to \emph{c} axis
(\textbf{\textcyr{\char209}}), in the plane parallel to \emph{c} axis
(\textbf{D}) \textcyr{\char232} chess-mate replacement (\textbf{E}).
The lattice parameters and calculated magnetic mome\textcolor{black}{nts
on the host iron atoms in }\textcolor{black}{\emph{$Fe$}}\textcolor{black}{{}
sublattice of }\textcolor{black}{\emph{$\alpha-FeSi_{2}$}}\textcolor{black}{{}
($Fe_{0}$) and on the substitutional iron atoms ($Fe_{I}$ and $Fe_{II}$)
obtained after full optimization of geometry are given in Table \ref{tab:Table 1}.
The geometry optimization results in t}he elongation of all cells
along \emph{c} axis and to the compression in the (\emph{ab}) plane
which are most pronounced for the \textbf{C} and \textbf{E} alloys.

\begin{widetext}

\begin{table}[H]
\caption{\textcolor{black}{\emph{(Color online)}} The structures of some of
ordered alloys, the optimized lattice parameters and the calculated
magnetic moments; the colors encode: \emph{$Si$} atoms by blue, host
$Fe_{0}$ atoms by grey, the substitutional $Fe_{I}$ and $Fe_{II}$
atoms by black and green, correspondingly. \label{tab:Table 1}}

\includegraphics[scale=0.6]{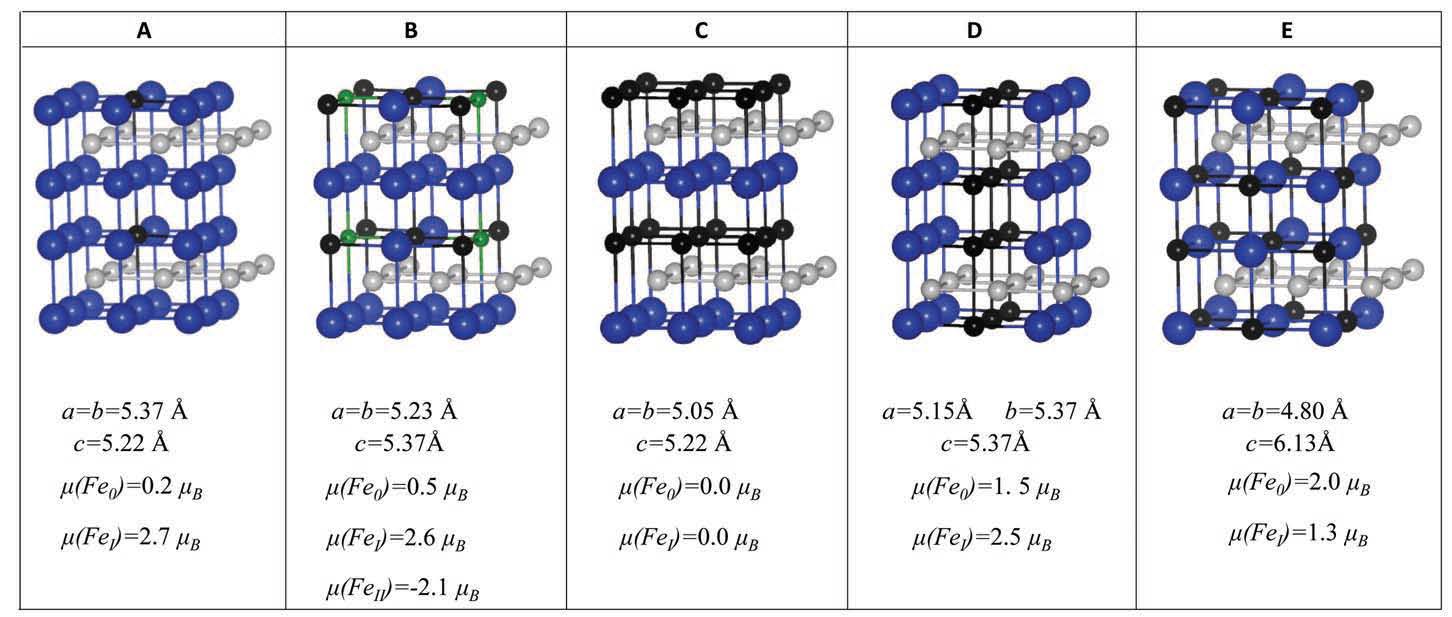}
\end{table}

\end{widetext}

\begin{figure}
\begin{tabular}{c}
\includegraphics[scale=0.3]{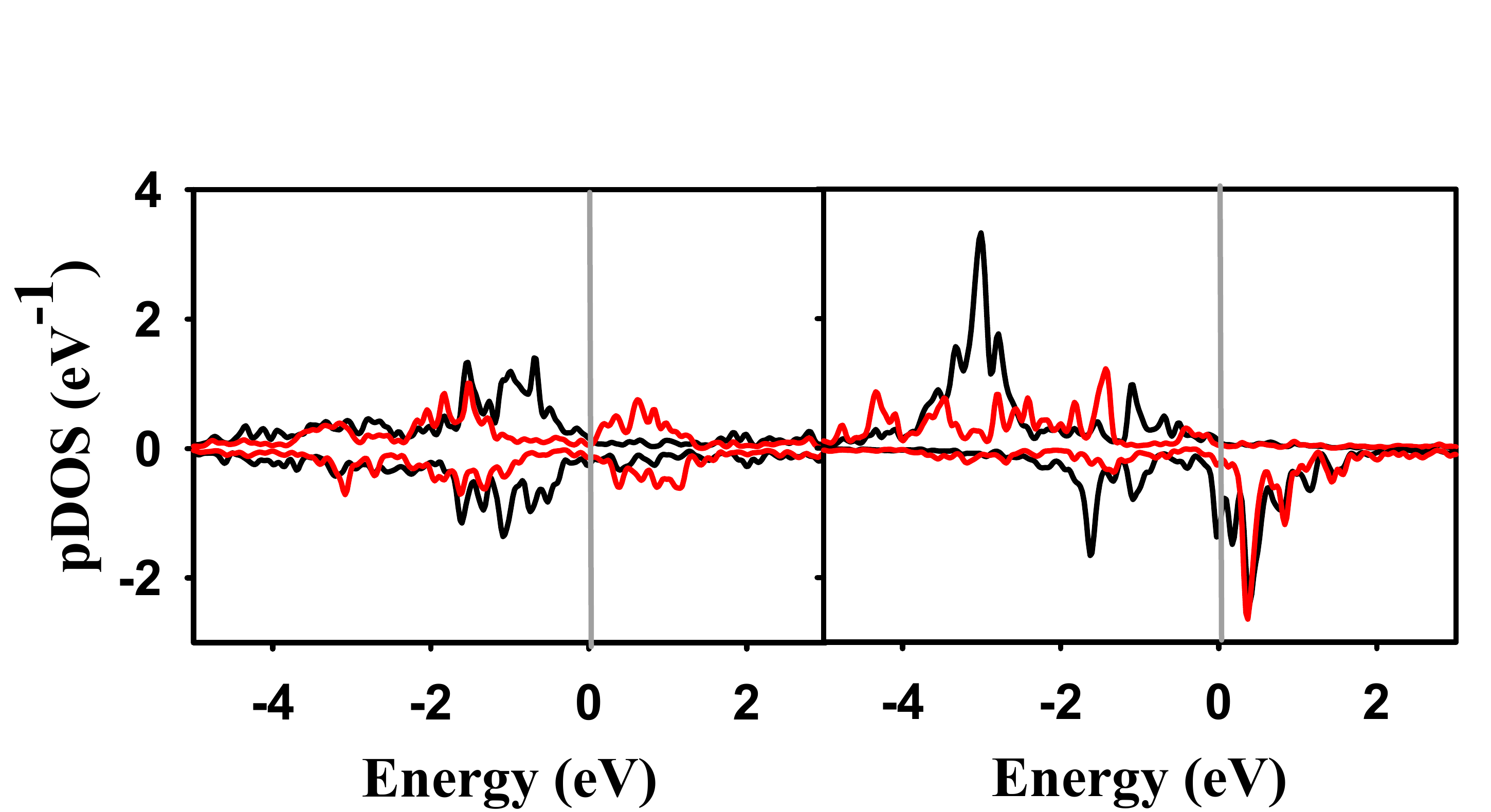}\tabularnewline
$\mu(Fe_{0})=$0.2$\mu_{B}$ $\qquad\quad\qquad\mu(Fe{}_{I})=$2.7$\mu_{B}$\tabularnewline
\end{tabular}

\caption{\emph{(Color online) }pDOS for host $Fe_{0}$ (left) anf substitutional
$Fe_{I}$ (right) in the alloy \textbf{A}. Black line shows $t_{2g}$
states, red line shows $e_{g}$ states. Zero on the energy axis is
the Fermi energy \label{fig:Fig.2}. }

\end{figure}

The substitution of one \emph{$Si$} atom by iron (\textbf{A})\textcolor{black}{{}
results in the appearance of the} large magnetic moments ($\mu(Fe{}_{I})=$2.7$\mu_{B}$)
on the substitutional $Fe_{I}$ atom. Alhough the alloy \textbf{A}
is ordered the obtained result is coincides with the\textcolor{black}{{}
result obtained in CPA \cite{key-16} for a random alloy. The value
of the magnetic moment and pDOS on the substitutional }\textcolor{black}{\emph{Fe}}\textcolor{black}{{}
atom are in a good agreement with the ones from Ref.\cite{key-16}}.
The general feature of both DOS is the sharp peak at the energy $\sim-3$eV,
which originates from the minority $t_{2g}$ state of \emph{$d$}-electrons
(Fig. \ref{fig:Fig.2}).

A further increase of substitutional \emph{$Fe$}\textcolor{black}{{}
concentration leads to }the non-trivial results that clearly illustrate
the dependence of \emph{$Fe$} magnetic moments on the local environment.
As seen from the Table \ref{tab:Table 1}, the substitution of three
\emph{$Si$} atoms by \emph{$Fe$} ones (alloy \textbf{B)} results
in the appearence of the \emph{ferri}magnetic state: the substitutional
$Fe_{I}$ and $Fe_{II}$ atoms become inequivalent: they acquire large
magnetic moments, which are not equal to each other, and directed
into \emph{opposite} directions. The absolute values of magnetic moments
are close to the ones in the alloy \textbf{A}.\textcolor{red}{{} }\textcolor{black}{The
alloy }\textbf{\textcolor{black}{\textcyr{\char194}}}\textcolor{black}{{}
presents only one of possible ways to order three substitutional }\textcolor{black}{\emph{Fe}}\textcolor{black}{{}
atoms in the supercell. Other nonequivalent ordering of the substitutional
$Fe$ atoms are shown in} Table \ref{tab:Table 2}. \textcolor{black}{Our}\textcolor{black}{\emph{
ab initio}}\textcolor{black}{{} calculations }show that the type of
the magnetic structure, ferrimagnetic or ferromagnetic, is determined
by the spatial arrangement of substitutional \emph{Fe} atoms. \textcolor{black}{Indeed,
the first two alloys in Table \ref{tab:Table 2} are ferrimagnetic,
and the last three are ferromagnetic}. The same dependence of the
iron magnetic moments on the spatial arrangement (and\textcolor{black}{{}
hence }on the the local environment) arises for the alloys with four
substitutional \emph{Fe} atoms on \emph{Si} sites (\textbf{C, D, E
}in Table \ref{tab:Table 1}). The alloy \textbf{\textcyr{\char209}}
and \emph{$\alpha-FeSi_{2}$} are non-magnetic, while the magnetic
moments in the alloys \textbf{D} and \textbf{E} appear on the substitutional
$Fe_{I}$ and on the host $Fe_{0}$ atoms. \textcolor{black}{The pDOSes
of substitutional $Fe_{I}$ in alloys }\textbf{\textcolor{black}{B,
D, E}}\textcolor{black}{{} are similar to ones in alloy }\textbf{\textcolor{black}{A}}\textcolor{black}{,
pDOS of $Fe_{II}$ in alloy }\textbf{\textcolor{black}{B}}\textcolor{black}{{}
atom is mirror-symmetric to pDOS of $Fe_{I}$. Notice that the} $t_{2g}$
states\textcolor{black}{{} form peak in pDOS of }substitutional $Fe_{I}$
\textcolor{black}{atom when the latter has magnetic moment while the
pDOS of $Fe_{I}$ }\textcolor{black}{\emph{d}}\textcolor{black}{-electrons
in the non-magnetic alloy }\textbf{\textcolor{black}{C}}\textcolor{black}{{}
is }similar to the one for the \emph{Fe} atom in \emph{$\alpha-FeSi_{2}$}(Fig.
\ref{fig:1a}b): $t_{2g}$ and $e_{g}$ electron states are delocalized
in the wide energy range.

Thus, our \emph{ab initio} calculations confirm only part of the conclusions,
derived from the local environment models \cite{key-18,key-19}: the
ferromagnetism arises with an increase of the \emph{Fe} concentration
indeed, but the types of the magnetic structure of the ordered $Fe_{1+x}Si_{2-x}$
alloys, which we obtain, are essentially different even at the same
concentration of substitutional \emph{Fe} atoms (Table \ref{tab:Table 1}
and Table \ref{tab:Table 2}): the magnetic moments on \emph{Fe} atoms
are determined by the composition and the configuration of its local
environment. These findings motivate us to investigate the role played
by the different local environment on the magnetic moments formation
in $Fe-Si$ alloys more carefully in the framework of the multiorbital
model suggested in \cite{key-21} and briefly outlined in Sec. II\textbf{.
}As was pointed out in \cite{key-21} the crucial role in the magnetic
structure formation in iron silicides is played by both nearest and
\emph{next}-nearest local environment. Both are taken into account
in a model calculations.

\begin{widetext}

\begin{table}[H]
\caption{\emph{(Color online) }The ordered alloys with three substitutional
$Fe$ atom at the \emph{$Si$} sites.\emph{ $Si$ }atoms are shown
by blue balls, host $Fe_{0}$ by grey, substitutional $Fe_{I}$ and
$Fe_{II}$ atoms are shown by black and green balls, correspondingly.\label{tab:Table 2}}

\includegraphics[scale=0.6]{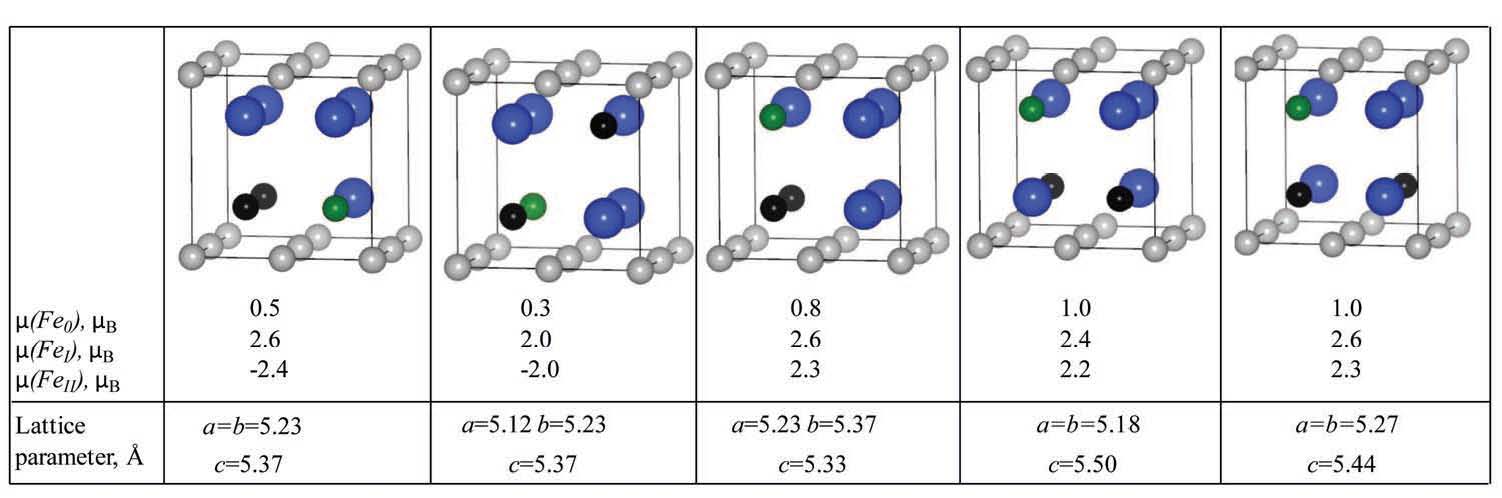}
\end{table}

\end{widetext}

\subsection{The model calculations}

In this subsection we describe the results of model calculations for
the stochiometric \emph{$\alpha-FeSi_{2}$ }and its ordered \emph{Fe}-rich
alloys\emph{ }$Fe_{1+x}Si_{2-x}$(\textbf{B}, \textbf{C} and \textbf{D}
in Table \ref{tab:Table 1}). In all model calculations we have used
the following parameters (see Sec.II): Hubbard $U=1$, \emph{i.e.}
all other parameters are given in units of \emph{U}; $J=0.4,\:J'=0.05,\:\varepsilon_{Si}=6,\:\varepsilon_{Fe}=0$.
In the general case there are five hopping parameters: $t_{1}$ (\emph{Fe-Fe})
and $t_{2}$ (\emph{Fe-Si}) between the nearest neighbors (NN); $t_{3}$
(\emph{Fe-Fe}), $t_{4}$ (\emph{Fe-Si}) between next-nearest neighbors
(NNN), and $t_{5}$ for \emph{Si \textendash Si} hoppings. The relation
$t_{\pi}=\frac{1}{3}t_{\sigma}$ for NN and $t_{\pi}=\frac{1}{2}t_{\sigma}$for
NNN was kept in all model calculations; for this reason further everywhere
we will use $t_{\sigma}\equiv t$. The values for these hopping parameters
are found from the requirement that after achieving self-consistency
in both the model and the \emph{ab initio} calculations (GGA), the
\emph{d}-DOS and magnetic moments on \emph{Fe} atoms have to be as
close to each other as possible. The best fit of the model magnetic
moments and DOS to the\emph{ ab initio }ones can be achieved only
when the hopping integrals are positive for the NN and negative for
NNN.\textcolor{black}{{} Along all model calculations we used equlibrium
lattice parameter, obtained from the }\textcolor{black}{\emph{ab initio}}\textcolor{black}{{}
calculation (see Table \ref{tab:Table 1})}\textcolor{red}{.} We also
take in\textcolor{black}{to account that t}he values of hopping integrals
should correlate with the distance between neighbors in all ordered
alloys and in \emph{$\alpha-FeSi_{2}$}. \textcolor{black}{The values
of hopping parameters which provide the best fit are shown in the
Table \ref{tab:Table 3}.}

\begin{table}
\caption{The distances $d$ (Å) between nearest neighbors (NN) and next-nearest
neighbors (NNN) and the values of hopping integrals \emph{$t$,} which
provide the best fit of the model charge densities to GGA-DFT ones.\label{tab:Table 3}}

\begin{tabular}{|c|c|c|c|c|c|c|c|c|}
\hline
 & \multicolumn{2}{c|}{\emph{$\alpha-FeSi_{2}$} } & \multicolumn{2}{c|}{\textbf{B}} & \multicolumn{2}{c|}{\textbf{C}} & \multicolumn{2}{c|}{\textbf{D}}\tabularnewline
\hline
 & d & \emph{t} & d & \emph{t} & d & \emph{t} & d & \emph{t}\tabularnewline
\hline
Fe-Si (NN) & 2.36 & 1.0 & 2.38 & 0.95 & 2.37 & 1.0 & 2.39 & 0.95\tabularnewline
\hline
Fe-Fe (NN)  & - & - & 2.40 & 0.9 & 2.44 & 0.85 & 2.43 & 0.85\tabularnewline
\hline
Fe-Si (NNN)  & - & - & 2.62 & -0.55 & 2.24 & -0.8 & 2.56 & -0.45\tabularnewline
\hline
\multirow{2}{*}{Fe-Fe (NNN) } & \multirow{2}{*}{2.70} & \multirow{2}{*}{-0.65} & \multirow{2}{*}{2.60} & \multirow{2}{*}{-0.70} & \multirow{2}{*}{2.53} & \multirow{2}{*}{-0.75} & 2.56 & -0.72\tabularnewline
 &  &  &  &  &  &  & \textcolor{black}{2.78} & -0.60\tabularnewline
\hline
Si-Si (NN)  & 2.34 & \multicolumn{1}{c|}{2.0} & 2.39 & 2.0 & 2.53 & 1.5 & 2.41 & 2.0\tabularnewline
\hline
Si-Si (NNN) & 2.80 & 1.0 & 2.61 & 1.5 & - & - & 2.78 & 1.0\tabularnewline
\hline
\end{tabular}
\end{table}

\subsubsection{$\alpha-FeSi_{2}$}

We begin with the stoichiometric \emph{$\alpha-FeSi_{2}$}(Fig. \ref{fig:1a}a).
It has the tetragonal lattice with the space group $P4/mm$. Each
of\emph{ $Fe$} atom in the \emph{$\alpha-FeSi_{2}$} has only \emph{$Si$}
atoms in the nearest local environment and only \emph{Fe} atoms as
the second neighbors, therefore, there are three hoppings integrals:
between NN \emph{$Fe-Si$} ($t_{2}$), between NNN \emph{$Fe-Fe$}
($t_{3}$) and between \emph{$Si-Si$} ($t_{5}$). These parameters
were used for fitting the model\emph{ d}-DOS and the magnetic moments
on \emph{$Fe$} atoms to the \emph{ab initio} ones. The values of
$t_{2}$, $t_{3}$and $t_{5}$ parameters which provide the best fitting
are shown in the Table \ref{tab:Table 3}. The model and GGA \emph{Fe
$d$}- population numbers for stoichiometric \emph{$\alpha-FeSi_{2}$
}and corresponding partial DOS of \emph{Fe} \emph{d}-electrons are
compared in Table \ref{tab:Table 4}. The accuracy of the statement
that the model reflects the properties of real compounds and qualitatively
the features of \emph{ab initio} pDOS at this set of parameters is
seen from the Table \ref{tab:Table 4}.

\begin{figure}
\includegraphics[scale=0.4]{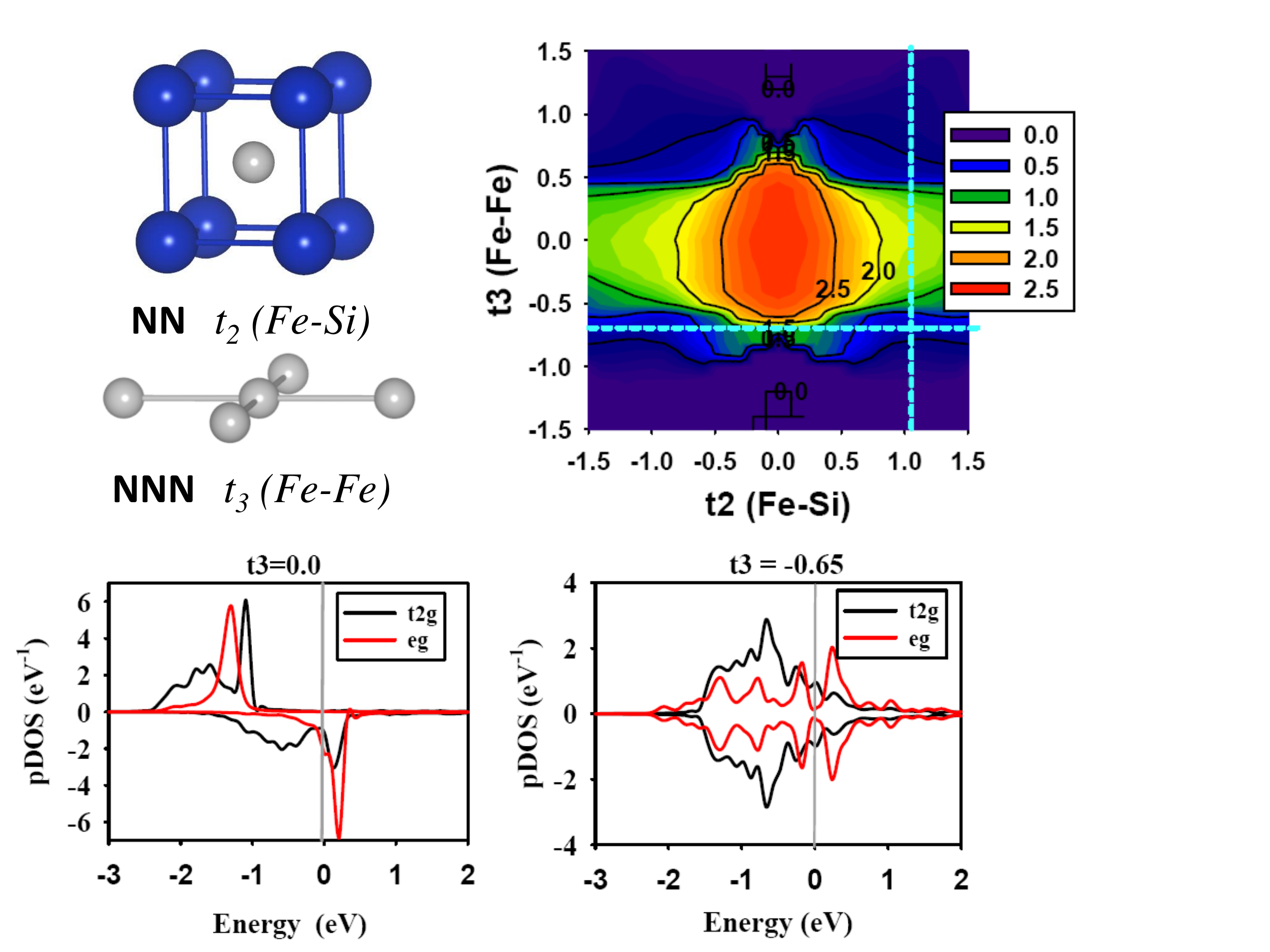}

\caption{\emph{(Color online) }Top panel: Nearest and next-nearest neighbors
of $Fe_{0}$ with corresponding hopping integrals (\emph{Fe }and \emph{Si}
atoms are shown by grey and blue balls correspondingly) and the $t_{2}-t_{3}$-map
of magnetic moments; the blue lines show the values of hopping integrals
$t_{2}$and $t_{3}$ from Table \ref{tab:Table 4}. Bottom panel:
Model pDOS for hopping integral $t_{3}=0.0$ (left) and $t_{3}=-0.65$
(right). Hopping integral $t_{2}$=1.0 \label{fig:Fig.4}}
\end{figure}

\begin{table}
\caption{The comparison of orbital population numbers ($n_{\uparrow}^{d}$,
$n_{\downarrow}^{d}$), magnetic moments ($\mu$) and the number of
electrons ($N_{el}$) for \emph{$\alpha-FeSi_{2}$} in the model with
GGA-DFT ones. The \emph{ab intio} (blue lines) and the model (black
lines) pDOS of \emph{Fe d}-electrons (left: $t_{2g}$ -electrons,
right: $e_{g}$-electrons) in \emph{$\alpha-FeSi_{2}$} are compared
in the figure under the Table.\label{tab:Table 4}.}

\begin{tabular}{|c|c|c|c|c|c|c|}
\hline
 & \multicolumn{3}{c|}{VASP} & \multicolumn{3}{c|}{Model}\tabularnewline
\hline
 & $n_{\uparrow}^{d}$ & $n_{\downarrow}^{d}$ & \begin{turn}{90}
\end{turn} & $n_{\uparrow}^{d}$ & $n_{\downarrow}^{d}$ & \tabularnewline
\hline
$d_{xy}$ & 0.77 & 0.76 & $\mu=0.1\mu_{B}$ & 0.67 & 0.66 & $\mu=0.2\mu$\tabularnewline
$d_{xz}$ & 0.72 & 0.71 & $N_{el}=6.6$ & 0.70 & 0.68 & $N_{el}=6.6$\tabularnewline
$d_{yz}$ & 0.72 & 0.71 &  & 0.79 & 0.68 & \tabularnewline
$d_{x^{2}-y^{2}}$ & 0.58 & 0.55 &  & 0.63 & 0.61 & \tabularnewline
$d_{z^{2}}$ & 0.67 & 0.63 &  & 0.70 & 0.60 & \tabularnewline
\hline
\end{tabular}

\includegraphics[scale=0.25]{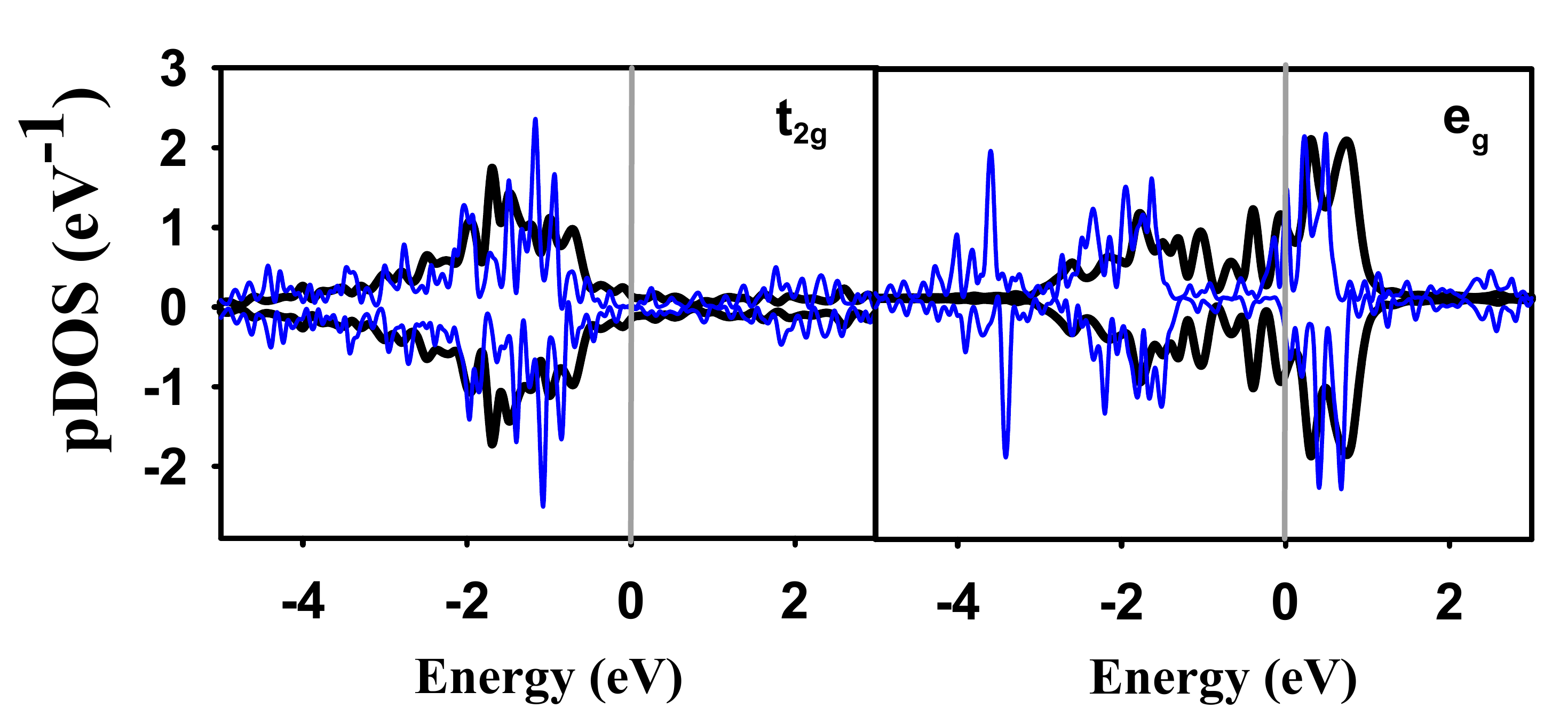}
\end{table}

In order to understand the effect of NN and NNN neighbors in the local
environment on the magnetic moment (MM) formation we calculaled the
dependence of the MMs on the hopping integrals $t_{2}$ (NN\emph{
Fe - Si}) and $t_{3}$ (NNN $Fe-Fe$). The map of the magnetic moment
dependences on the hopping integrals $t_{2}$ and $t_{3}$ is shown
in the top panel of Fig. \ref{fig:Fig.4}. As seen the crucial role
in the MM formation is played by hoppings between NNN \emph{Fe - Fe}
($t_{3}$). Indeed, with $\left|t_{3}\right|>0.6$ the experimentally
existing nonmagnetic state is stable, a decrease of $\left|t_{3}\right|$
leads to the transition into ferromagnetic state. \textcolor{black}{Furthermore,
the boundaries between region with magnetic states and non magnetic
ones are very sharp} (Fig. \ref{fig:Fig.4}, top panel):\textcolor{black}{{}
the MM decreases till zero very fast as a function of hopping $t_{3}$
between iron atoms.} The hopping between NN \emph{Fe - Si} ($t_{2}$)
has effect only on the magnitude of the MM in the ferromagnetic region.
The mechanism of ferromagnetism destruction with hopping $t_{3}$
is clearly seen from the bottom panel of Fig. \ref{fig:Fig.4}. Switching
off the hopping between NNN \emph{Fe \textendash{} Fe} ($t_{3}=0$)
makes the $d$-bands atom-like with the slight smearing. An increase
of the $t_{3}$ hopping leads to a delocalization of these atom-like
\emph{d}-bands and destruction the magnetism. Hence, an increase of
the distance between NNN (or, a decrease the hopping integral $t_{3}$)
would results in the transition from nonmagnetic phase to magnetic
one. \textcolor{black}{This conclusion from the analysis of the model
is confirmed by the }\textcolor{black}{\emph{ab initio}}\textcolor{black}{{}
calculation: the increase of the lattice parameters }\textcolor{black}{\emph{a}}\textcolor{black}{{}
and }\textcolor{black}{\emph{b}}\textcolor{black}{{} of }\textcolor{black}{\emph{$\alpha-FeSi_{2}$}}\textcolor{black}{{}
(or the distance NNN }\textcolor{black}{\emph{$Fe-Fe$}}\textcolor{black}{)
by 7\% ($a=b=$2.9Å, $\lyxmathsym{\textcyr{\char241}}=$5.13 Å) causes
formation of MMs $\mu=0.6\mu_{B}$ on the }\textcolor{black}{\emph{Fe}}\textcolor{black}{{}
atoms.}\textbf{\textcolor{black}{{} }}\textcolor{black}{Thus, it is
rather the hopping integral between the NNN }\textcolor{black}{\emph{$Fe-Fe$
}}\textcolor{black}{atoms, not the NN $Fe-Si$ hopping, determines
the existence of magnetic or nonmagnetic state in }\textcolor{black}{\emph{$\alpha-FeSi_{2}$}}\textcolor{black}{,
because the NN of }\textcolor{black}{\emph{Fe}}\textcolor{black}{{}
atom consist of }\textcolor{black}{\emph{Si}}\textcolor{black}{{} atoms
in the both cases. }

\subsubsection{Fe-rich alloys }

To emphasize the importance of the NNN in the MM formation on iron
atoms we consider the alloys \textbf{C} and \textbf{D} from Table
\ref{tab:Table 1}. These alloys reveal essentially different magnetic
behavior at the same concentration but different spatial arrangements
of the substitutional \emph{$Fe$} atoms.

\begin{widetext}

\begin{figure}
\includegraphics[scale=0.6]{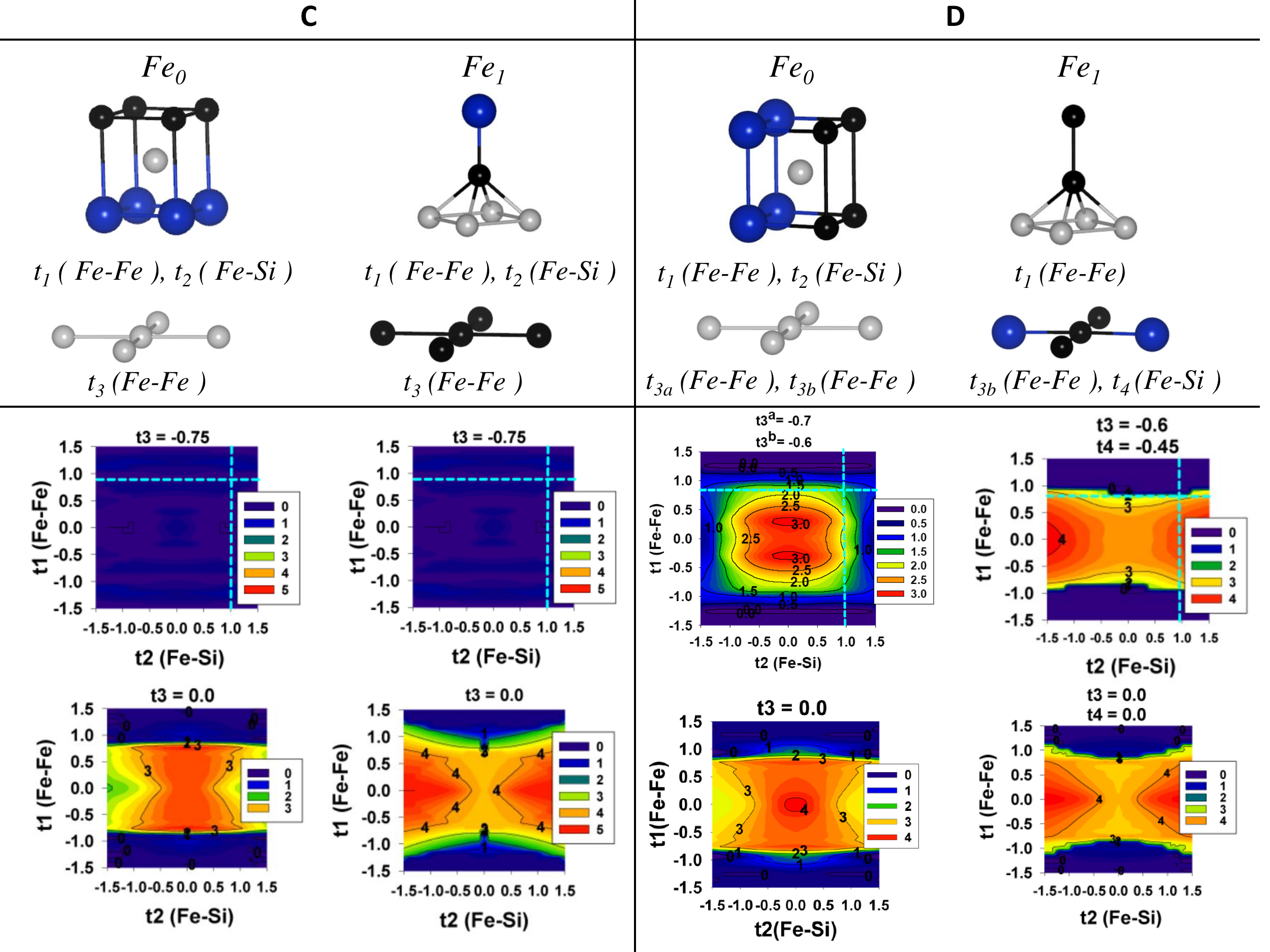}

\caption{(\emph{Color online}) Alloy \textbf{C} and\textbf{ D}. \emph{Top panel:}
NN and NNN environment of iron atoms. \emph{$Si$} atoms are shown
by blue balls, grey and black balls stand for $Fe_{0}$ and substitutional
$Fe_{I}$ atoms, correspondingly. \emph{Middle panel:} Dependence
of the MMs on hopping integrals $t_{1}$ and $t_{2}$ (hopping integrals
$t_{3}$ and $t_{4}$ are switch on). \emph{Bottom panel:} Dependence
of MMs on hopping integrals $t_{1}$ and $t_{2}$ (hopping integrals
$t_{3}$ and $t_{4}$ are switch off). Blue lines show the values
of hopping integrals $t_{1}=0.9$ and $t_{2}=1.0$ for alloy \textbf{C}
and $t_{1}=0.85$ and $t_{2}=0.95$ for alloy \textbf{D} (Table \ref{tab:Table 3});
these values provide the best fitting to the\emph{ ab initio} charge
density. \label{fig:Fig.5}}
\end{figure}

\end{widetext}

\textcolor{black}{As it follows from}\textcolor{black}{\emph{ ab initio}}\textcolor{black}{{}
calculation, the ordered alloy }\textbf{\textcolor{black}{D}}\textcolor{black}{{}
reveals ferromagnetism, whereas the alloy }\textbf{\textcolor{black}{\textcyr{\char209}}}\textcolor{black}{{}
remains nonmagnetic (Table \ref{tab:Table 1}).} These ordered alloys
have two nonequivalent \emph{$Fe$} atoms: $Fe_{0}$ is the host iron
atom in the iron sublattice of \emph{$\alpha-FeSi_{2}$} and $Fe_{I}$
is the substitutional \emph{Fe} atoms in the \emph{Si} sublattice.\textcolor{red}{{}
}\textcolor{black}{Different spatial arragenment of the substitutional
}\textcolor{black}{\emph{Fe}}\textcolor{black}{{} atoms results in the
different environment of the host and sustitutional }\textcolor{black}{\emph{Fe}}\textcolor{black}{{}
atoms in alloys }\textbf{\textcolor{black}{D}}\textcolor{black}{{} and
}\textbf{\textcolor{black}{C}}\textcolor{black}{.} \textcolor{black}{These
environments are shown in Fig. \ref{fig:Fig.5} (top panel).}\textcolor{red}{{}
}There is the important difference in the NNN environment of $Fe_{0}$
and $Fe_{1}$ in \textbf{C} and \textbf{D} alloys. In the \textbf{\textcyr{\char209}}
alloy both $Fe_{0}$ and $Fe_{I}$ atoms have four \emph{$Fe$} atoms
along crystallographic axes \textbf{a} and \textbf{b} as NNN \textcolor{black}{at
the same distances $R=2.53$Å (Table \ref{tab:Table 3}). The host
$Fe_{0}$ in the alloy }\textbf{\textcolor{black}{D}}\textcolor{black}{{}
also has four NNN $Fe$ atoms, but at different distances: two $Fe$
neighbors along axis }\textbf{\textcolor{black}{\emph{a}}}\textcolor{black}{\emph{
}}\textcolor{black}{with the distance $R=2.56$Å and two ones along
axis }\textbf{\textcolor{black}{\emph{b}}}\textcolor{black}{\emph{
}}\textcolor{black}{with $R=2.78$Å. These inequal distances arise
due to the different symmetry of crystal lattices: the }\textbf{\textcolor{black}{C}}\textcolor{black}{{}
lattice is tetragonal with the space group $P4mm$, while the }\textbf{\textcolor{black}{D}}\textcolor{black}{{}
one is orthorhombic with space group $Pmmm$. Thus, the distortions
of the underlying tetragonal lattice of }\textcolor{black}{\emph{$\alpha-FeSi_{2}$}}\textcolor{black}{{}
arising in these alloys are different. Notice, that the distances
between the NN }\textcolor{black}{\emph{$Fe-Si$}}\textcolor{black}{{}
and the NN }\textcolor{black}{\emph{$Fe-Fe$}}\textcolor{black}{{} in
both alloys are the same. Besides, the atom $Fe_{I}$ has only two
NNN }\textcolor{black}{\emph{$Fe$}}\textcolor{black}{{} atoms at the
distance $R=2.78$Å in the alloy }\textbf{\textcolor{black}{D}}\textcolor{black}{.
This distance is larger than the corresponding one in the alloy }\textbf{\textcolor{black}{C}}\textcolor{black}{.
Therefore, we are forced to introduce in the alloy }\textbf{\textcolor{black}{D}}\textcolor{black}{{}
two hopping integrals for the short $t_{3}^{a}$ and long $t_{3}^{b}$}\textbf{\textcolor{black}{{}
}}\textcolor{black}{distances between NNN Fe -Fe, while only one hopping
integral $t_{3}$ is required for the description of the alloy}\textbf{\textcolor{black}{{}
C}}\textcolor{black}{. The values of hopping integrals providing the
best fitting to the }\textcolor{black}{\emph{ab initio}}\textcolor{black}{{}
calculation are given in Table \ref{tab:Table 3}. The Hartree-Fock
self-consistent MMs generated by the model at these values of hopping
parameters are$\mu^{mod}(Fe_{0})=\mu^{mod}(Fe_{1})=0$ in alloy }\textbf{\textcolor{black}{C}}\textcolor{black}{{}
and $\mu^{mod}(Fe_{0})=1.4\mu_{B},\:\mu^{mod}(Fe_{I})=2.5\mu_{B}$
in alloy }\textbf{\textcolor{black}{D}}\textcolor{black}{.}

Let us compare the dependences of the \emph{Fe} magnetic moments on
the NN hopping integral $t_{1}$ and $t_{2}$ at fixed values of NNN
hopping integral $t_{3}$ and $t_{4}$, shown at the middle panel
of Fig.4. The range of the magnetic moments existence on both $Fe_{0}$
and $Fe_{I}$ atoms in the alloy \textbf{D} is restricted by the values
of $\left|t_{1}\right|<1$. The magnetic state with moments close
to \emph{ab initio} values ($\mu(Fe_{0})=1.5\mu_{B},\:\mu(Fe_{I})=2.5\mu_{B}$)
is on the narrow boundary between ferro- and paramagnetic phases.
In the alloy \textbf{\textcyr{\char209}} the nonmagnetic state is
stable in all range of the hoppings between NN $t_{1}$ and $t_{2}$.
Namely the circumstance that the magnetic moments are close to the
instability line make them very sensitive to changes of the NNN hoppings.
Indeed, which of solutions, magnetic or non-magnetic, will arise,
is controlled by the value of hopping integral $t_{3}$: $t_{3}^{\boldsymbol{C}}(Fe-Fe)=-0.75$
leads to formation of the paramagnetic state in the alloy \textbf{\textcyr{\char209}},
whereas a decrease of $t_{3}$ in the alloy \textbf{D}, $t_{3}^{\boldsymbol{D}}(Fe-Fe)=-0.60$
gives birth to a ferromagnetic state in the alloy \textbf{D}. The
increase of $\left|t_{3}^{\boldsymbol{C}}(Fe-Fe)\right|$ compared
to $\left|t_{3}^{\boldsymbol{D}}(Fe-Fe)\right|$ occurs due to shorter
distance between \emph{Fe} atoms in the NNN environment in alloy \textbf{\textcyr{\char209}}
(Table \ref{tab:Table 3}). Moreover, a decrease of $\left|t_{3}\right|$
results in appearance of magnetic moments on both \emph{Fe} atoms
in the alloy \textbf{C;} at $t_{3}^{\boldsymbol{C}}=0.0$ the map
of magnetic moments in the alloy \textbf{C }becomes similar to one
for the alloy \textbf{D} (Fig. \ref{fig:Fig.5}, bottom panel). At
first glance one could expect that the formation of the Fe-Si bond
should destroy the moment on the Fe atom. However, the magnetic moments
on $Fe$ atoms happen to be much less sensitive to the hopping parameter
$t_{2}$ between NN \emph{Fe} and \emph{Si} atoms. Indeed, all the
$t_{1}-t_{2}$-maps for \emph{Fe} moments, calculated within this
model, are elongated along the axis $t_{2}$.

\begin{figure}
\includegraphics[scale=0.5]{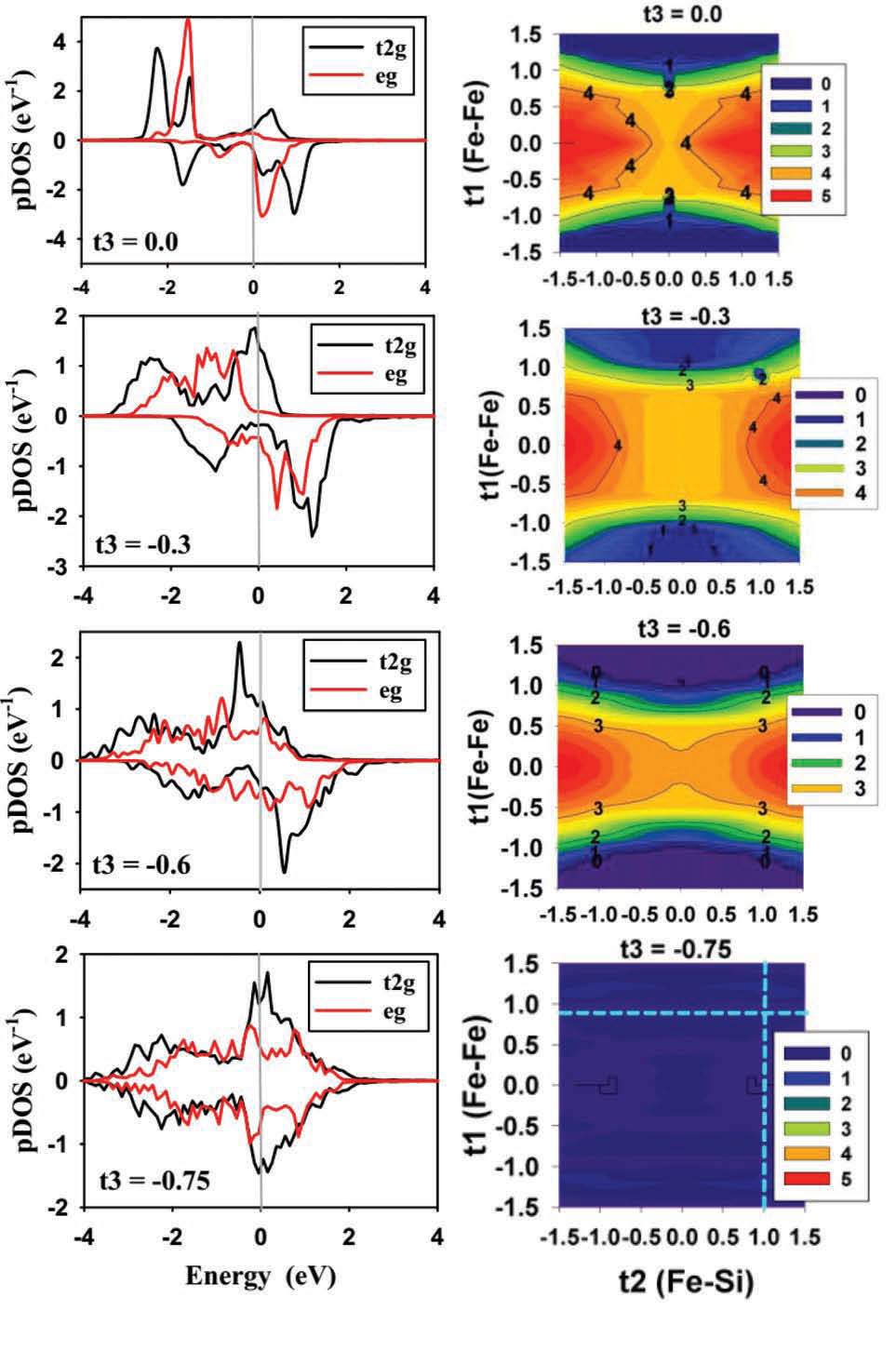}

\caption{(\emph{Color online}) The alloy \textbf{C}: Model pDOS (left panel)
and the map of the magnetic moments (right panel) for the different
values of hopping integral $t_{3}(Fe-Fe)$. The blue lines at the
last map show the values of hopping integrals $t_{1}=0.9$ and $t_{2}=1.0$
(Table \ref{tab:Table 3}), which provide the best fitting to the
\emph{ab initio} charge density. \label{fig:Fig.6}}
\end{figure}

The physics of the destruction of the magnetic moments on Fe atoms
can be interpreted from the point of view of the $d$-band formation.
The Fig. \ref{fig:Fig.6} illustrates this for the alloy \textbf{C}
via the evolution of the $Fe_{I}$\emph{ d}-electron pDOS and corresponding
magnetic-moment maps with the increasing of only hopping integral
$t_{3}$ at all other hopping integrals kept fixed. As seen, at first
steps of increase of $t_{3}$ a gradual smearing of initially (at
$t_{3}=0$) atom-like levels and a slight change of the map of magnetic
moments occurs. Then, similar to the case of \emph{$\alpha-FeSi_{2}$},
at $t_{3}=-0.75$ the abrupt destruction of the magnetic moments arises
and the difference between the minority and majority spin states in
pDOS disappears.

\begin{figure}
\includegraphics[scale=0.35]{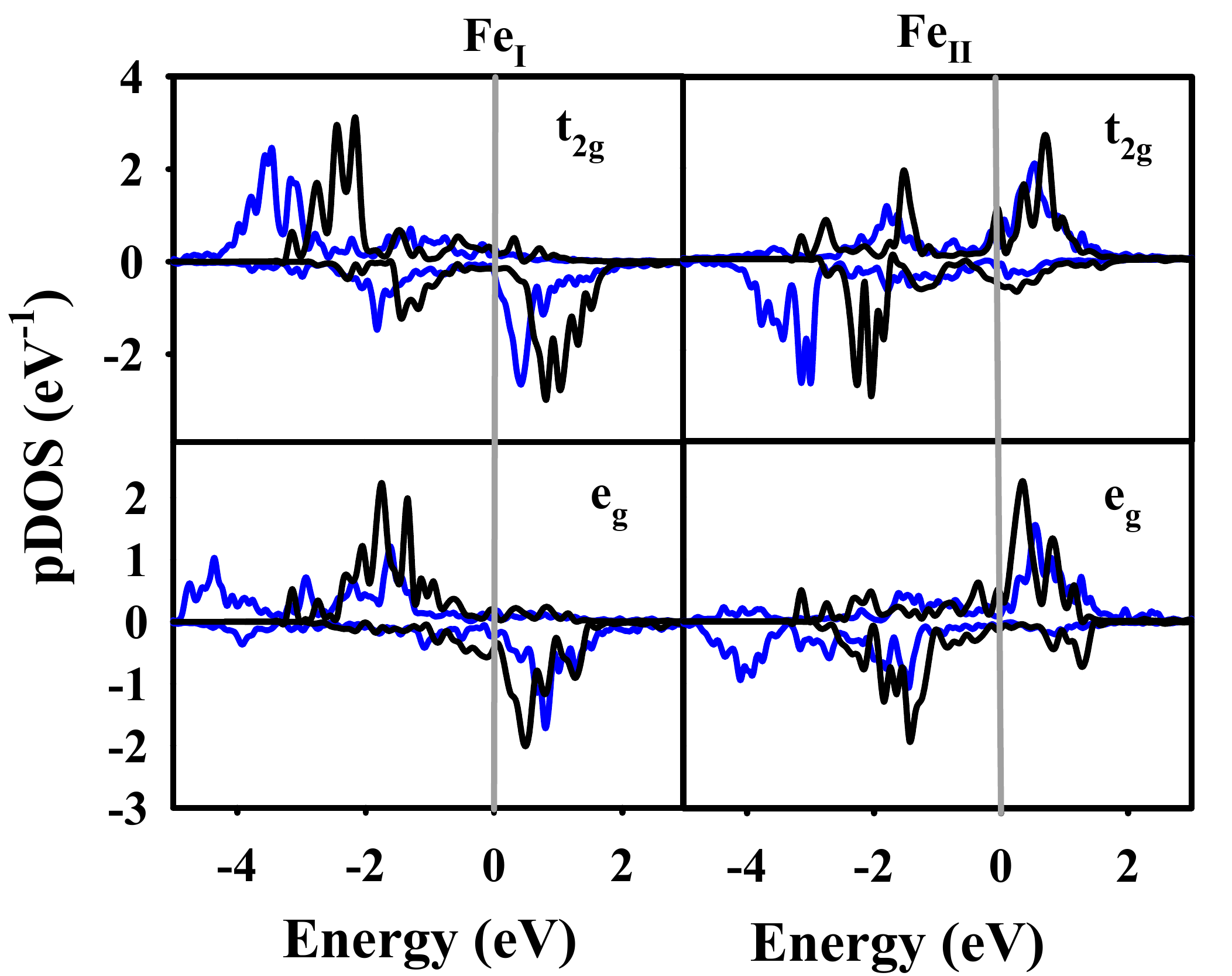}

\caption{(\emph{Color online}) The comparison of the \emph{ab initio} (blue
lines) and the model (black lines) pDOS of $Fe_{I}$\emph{ d}-electrons
(left panel) and $Fe_{II}$ \emph{d}-electrons (right panel) in the
alloy \textbf{B}. Top: $t_{2g}$-electrons, bottom: $e_{g}$-electrons\label{fig:Fig.7}}

\end{figure}

\begin{widetext}

\begin{figure}
\textcolor{black}{\includegraphics[scale=0.6]{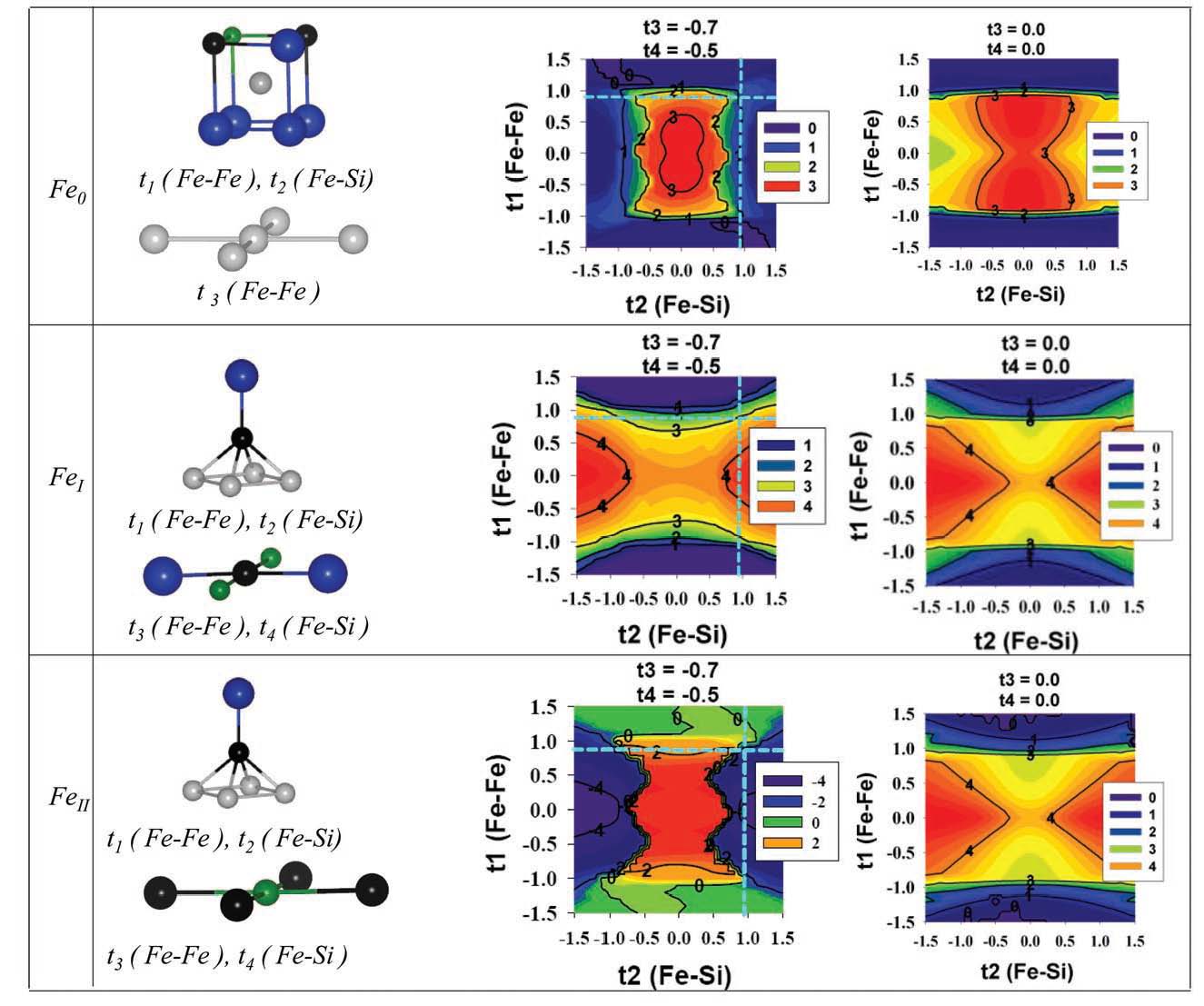}}

\caption{(\emph{Color online}) The alloy \textbf{B}. \emph{Top panel,} \emph{left:}
NN and NNN environment of $Fe_{0}$ atom; color encodings: \emph{Si}
are blue, $Fe_{0}$ are grey, substitutional $Fe_{I}$ and $Fe_{II}$
are black and green balls correspondingly;\emph{Top panel,} \emph{center:}
the dependence of $Fe_{0}$ MMs on hopping integrals $t_{1}$ and
$t_{2}$ (hopping iontegrals $t_{3}$ and $t_{4}$ are swich on);
\emph{Top panel,} \emph{right:} the dependence of MM on $Fe_{0}$
atom on the hopping integrals $t_{1}$ and $t_{2}$ (hopping integrals
$t_{3}$ and $t_{4}$ are swith off). \emph{Middle and bottom panels:}
the same for $Fe_{I}$ and $Fe_{II}$ atoms, correspondingly. \textcolor{black}{Blue
lines on the maps show the values of hopping integrals $t_{1}=0.9$
and $t_{2}=0.95$} (Table \ref{tab:Table 3}), which provide the best
fitting to the \emph{ab initio} charge density.\label{fig:Fig.8}}

\end{figure}

\end{widetext}

Let us now discuss the origin of the unusual ferrimagnetic state in
the type of the alloy \textbf{B }(Table \ref{tab:Table 1}), which
contains three substitutional \emph{$Fe$} atoms on the \emph{$Si$}
sites. The ordered alloy \textbf{B} has the tetragonal lattice with
space group $P4mm$. There are three non-equivalent \emph{$Fe$} atoms
in the unit cell: $Fe_{0}$ is the host iron sublattice of \emph{$\alpha-FeSi_{2}$},
$Fe_{I}$ and $Fe_{II}$ are the non-equivalent substitutional \emph{$Fe$}
atoms in the\emph{ $Si$} sublattice. In accordance with \emph{ab
initio} calculations, the absolute values of magnetic moments on $Fe_{I}$
and $Fe_{II}$ atoms are close to each other, but have \emph{opposite}
directions:\textcolor{black}{{} $\mu(Fe_{I})=2.3\mu_{B},\:\mu(Fe_{II})=-1.9\mu_{B}$}.
\textcolor{black}{The model MMs obtained for the values of hopping
integrals from Table \ref{tab:Table 3} are $\mu^{mod}(Fe_{I})=2.8\mu_{B}$,
$\mu^{mod}(Fe_{II})=-2.1\mu_{B}$, and $\mu^{mod}(Fe_{0})=0.7\mu_{B}$.
The specific feature of }\textcolor{black}{\emph{$Fe$}}\textcolor{black}{-pDOS
in the alloy }\textbf{\textcolor{black}{B}}\textcolor{black}{{} is that
the $Fe_{II}$ -pDOS is mirror-symmetric to the pDOS of $Fe_{I}$
atoms. This feature arises in both first-principle and model calculation.
The comparison of pDOSes for substitutional }\textcolor{black}{\emph{$Fe$
}}\textcolor{black}{atoms is shown in Fig. \ref{fig:Fig.7}. As in
previous cases we built the $t_{1}-t_{2}$-maps of MMs for three non-equivalent
}\textcolor{black}{\emph{Fe}}\textcolor{black}{{} atoms (Fig. \ref{fig:Fig.8}).
The bright illustration of the importance of NNN interactions is that
in spite of the fact that the NN local environment of substitutional
}\textcolor{black}{\emph{Fe}}\textcolor{black}{{} atoms is the same
(Fig. \ref{fig:Fig.8}, first column), they have completely different
maps of magnetic moments.}\textcolor{red}{{} }There is a wide range
(at $\left|t_{1}\right|\gtrsim0.5$) of the negative MMs in the map
for $Fe_{II}$ atom (Fig. \ref{fig:Fig.8}, bottom panel, middle column)
with the sharp boundary between positive and negative values of MMs,
\textcolor{black}{whereas in the same region of $Fe_{I}$ $t_{1}-t_{2}$-
map the MM remains positive. These distinctions occur due to different
number of Fe atoms in the NNN environment. }Indeed, switching off
the hoppings between NNN neighbors ($t_{3}$ and $t_{4}$) changes
the behavior of magnetic moments on $Fe_{II}$ atom: the region with
the negative moment disappears and the maps for $Fe_{I}$ and $Fe_{II}$
atoms became almost identical (cf. middle and bottom panel of Fig.
\ref{fig:Fig.8}, last column). \textcolor{black}{This numeric experiment
explicitely shows that the role played by the NNN local environment
is critically essential for the emergence of the $Fe$ atoms with
the opposite MMs and, correspondingly, for the development of the
ferrimagnetic state. }

\begin{widetext}

\begin{figure}
\includegraphics[bb=0bp 0bp 1200bp 379bp,scale=0.35]{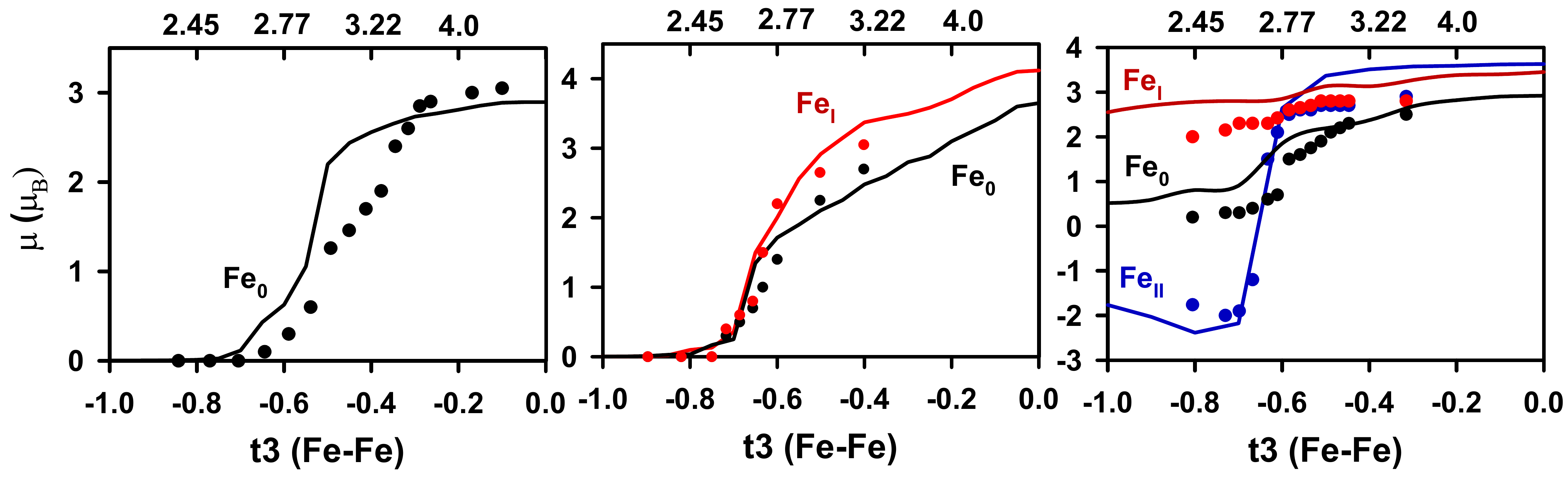}

\textcolor{black}{\emph{$\qquad\qquad\qquad\qquad\alpha-FeSi_{2}$}}
~$\qquad\qquad\qquad\qquad\qquad$ alloy \textbf{C} $\qquad\qquad\qquad\qquad$
$\qquad\qquad$ alloy \textbf{B}\textcolor{black}{\emph{ }}

\caption{(\emph{Color online}) The dependence of MMs on the host\emph{ $Fe_{0}$}
and the substitutional \emph{$Fe_{I}$ }and\emph{ $Fe_{II}$} atoms
on hopping $t_{3}$ in \textcolor{black}{\emph{$\alpha-FeSi_{2}$}},
the alloys \textbf{C} and \textbf{B }(from left to right). The solid
lines are the dependences, obtained in model calculations, the dots
show the MMs from \emph{ab initio} calculations at the distance R
between NNN Fe-Fe, according to (\ref{eq:3}). The scale for the distances
(in Å) is given on the tops of the figures.\label{fig:Fig.9}}
\end{figure}

\end{widetext}

\textcolor{black}{Thus, our analysis of the Hartree-Fock solutions
of the multiorbital models of iron silicides and supporting them first-principle
calculations allow to conclude that the decisive role in the destruction/formation
of the iron magnetic moments is played by the NNN local environment
or, more specifically, by the number of neighbors }\textcolor{black}{\emph{$Fe-Fe$}}\textcolor{black}{{}
and the by spacing between them. The results of our calculations show
that the previous statement \cite{key-18}-\cite{key-20}, that the
destruction of magnetic moments in the iron silicides is caused by
the increase of }\textcolor{black}{\emph{Si}}\textcolor{black}{{} atoms
in the NN environment is inaccurate. The obtained in our calculations
strong influence of NNN }\textcolor{black}{\emph{$Fe-Fe$}}\textcolor{black}{{}
couples is caused by the peculiarity of the}\textcolor{black}{\emph{
$\alpha-FeSi_{2}$}}\textcolor{black}{{} crystall structure, where the
the iron atoms form planes. }Since NNN \emph{$Fe-Fe$} are arranged
along crystallographic axes, the strong $\sigma$-bonds \textcolor{black}{between
}\textcolor{black}{\emph{Fe}}\textcolor{black}{{} atoms} are formed.\textcolor{red}{{}
}\textcolor{black}{So }in the alloy \textbf{C} and \emph{$\alpha-FeSi_{2}$}
which contains the iron (001) planes with the shorter distance between
\emph{$Fe-Fe$} than in the alloy \textbf{D}, these \emph{d}-bonds
result in the delocalization of the electrons and a decrease of the
MMs up till their destruction (Fig.\ref{fig:Fig.6}). At the same
time in the alloys \textbf{C} and \textbf{D }\emph{$Fe$} atoms have
the same number of \emph{$Si$} atoms in the NN environment and this
does not prohibit them to have different MMs. It is very instructive
to have a look from this point of view at the MM formation in the
alloy \textbf{A} where the substitutional \emph{Fe} atoms have maximal
MMs compared to the ones in all other alloys considered here. The
NNN environment of the substitutional $Fe_{I}$ atom in the alloy
\textbf{A} (Table \ref{tab:Table 1}, first column) consists of only
\emph{Si} atoms; the hoppings between \emph{$Fe-Fe$} which are responsible
for the destruction of moment are absent. This facts lead to the formation
of a large value of MM on this iron atom.

\textcolor{black}{In order to demonstrate the decisive role of the
}\textcolor{black}{\emph{$d-d$}}\textcolor{black}{-hopping integral
$t_{3}$ between NNN}\textcolor{black}{\emph{ $Fe-Fe$}}\textcolor{black}{{}
on the formation of the MM on }\textcolor{black}{\emph{Fe }}\textcolor{black}{atoms
we calculate the dependence of }\textcolor{black}{\emph{$Fe$}}\textcolor{black}{{}
MM on this hopping for }\textcolor{black}{\emph{$\alpha-FeSi_{2}$
}}\textcolor{black}{and the alloys }\textbf{\textcolor{black}{C}}\textcolor{black}{{}
and }\textbf{\textcolor{black}{B.}}\textcolor{black}{{} This dependence
is shown at Fig. \ref{fig:Fig.9}. As seen, the increase of $t_{3}$
in the alloy }\textbf{\textcolor{black}{C }}\textcolor{black}{and}\textbf{\textcolor{black}{{}
}}\textcolor{black}{in}\textcolor{black}{\emph{ $\alpha-FeSi_{2}$}}\textcolor{black}{{}
causes a destruction of the }\textcolor{black}{\emph{Fe}}\textcolor{black}{{}
MMs, whereas in the alloy }\textbf{\textcolor{black}{B}}\textcolor{black}{{}
the abrupt flip of the $Fe_{II}$ magnetic moment is occurred with
an increase $t_{3}$. The model results are confirmed by the }\textcolor{black}{\emph{ab
initio}}\textcolor{black}{{} calculations. Obviously, the hopping integral
$t_{3}$ changes its value with an increase of the spacing between
NNN $Fe-Fe$. Since the integral of the hopping matrix element contains
an overlap of the wave functions, we assume that it depends on the
distance }\textcolor{black}{\emph{R}}\textcolor{black}{{} between the
ions exponentially, }

\textcolor{black}{
\begin{equation}
t_{3}(R)=t_{3}^{max}exp(\gamma\Delta R),\label{eq:3}
\end{equation}
}

\textcolor{black}{where $t_{3}^{max}=t_{3}(R_{min})$ and $\Delta R=R-R_{min}$(Å).
Taking the values $t_{3}^{max}=-0.75$ and $R_{min}=2.53$ from Table
\ref{tab:Table 3} we have found the parameter $\gamma=-0.8926$$\mathrm{(\mathring{A}}^{-1}$).
Using Eq.(\ref{eq:3}) we obtained the distances }\textcolor{black}{\emph{R}}\textcolor{black}{{}
between NNN $Fe-Fe$ corresponding to the model parameters $t_{3}$.
Then the values of the MMs for the lattice parameters? corresponding
to these distances have been calculated within GGA-to-DFT. These values
are shown in Fig. \ref{fig:Fig.9} by dots. Remarkably, although the
only $t_{3}$ hopping was changed with the distance }\textcolor{black}{\emph{R}}\textcolor{black}{{}
in the model calculations (the values of the other hopping parameters
were kept fixed according to Table \ref{tab:Table 3}) we obtained
the good agreement between the model and the}\textcolor{black}{\emph{
ab initio}}\textcolor{black}{{} magnetic moments. This again proves
the significance of the NNN $Fe-Fe$ couplings for the MM formation.}

.

\subsection{\emph{Ab initio} calculation of \emph{the Si}-rich alloys.}

\textcolor{black}{Our model calculations lead to the conclusion that
the }\textcolor{black}{\emph{$Fe$}}\textcolor{black}{{} local MM formation
is controlled either by a decrease of the number of }\textcolor{black}{\emph{$Fe\lyxmathsym{\textendash}Fe$}}\textcolor{black}{{}
couples in }\textcolor{black}{\emph{$Fe$}}\textcolor{black}{{} layers
or by an increase of the distance between }\textcolor{black}{\emph{$Fe$}}\textcolor{black}{{}
atoms in pairs. Moreover, we can state that the increase of the cell\textquoteright s
magnetic moment with increase of}\textcolor{black}{\emph{ $x$}}\textcolor{black}{{}
in }\textcolor{black}{\emph{$Fe$}}\textcolor{black}{-rich alloys
$Fe_{1+x}Si_{2-x}$ is associated namely with the appearance of high-spin
}\textcolor{black}{\emph{$Fe$}}\textcolor{black}{{} species in the}\textcolor{black}{\emph{
$Si$}}\textcolor{black}{{} layers, which are surrounded mainly by the
}\textcolor{black}{\emph{$Si$}}\textcolor{black}{{} atoms. However,
these conditions can be fulfilled also by an increase of the }\textcolor{black}{\emph{$Si$}}\textcolor{black}{{}
concentration. To make sure that this unexpected conclusion derived
from the model is correct we carried out the }\textcolor{black}{\emph{ab
initio}}\textcolor{black}{{} GGA calculation of }\textcolor{black}{\emph{$Fe$}}\textcolor{black}{{}
magnetic moments for the }\textcolor{black}{\emph{$Si$}}\textcolor{black}{-rich
ordered alloys $Fe_{1-x}Si_{2+x}$. The alloy's structures must satisfy
the conditions listed above. By adding $Si$ atoms into the iron planes
we can decrease the number of the }\textcolor{black}{\emph{$Fe-Fe$}}\textcolor{black}{{}
couples. Besides, the substitutional }\textcolor{black}{\emph{$Si$}}\textcolor{black}{{}
atoms increase the spacing between the}\textcolor{black}{\emph{ $Fe$}}\textcolor{black}{{}
atoms.}

\begin{widetext}

\begin{figure}
\includegraphics[scale=0.65]{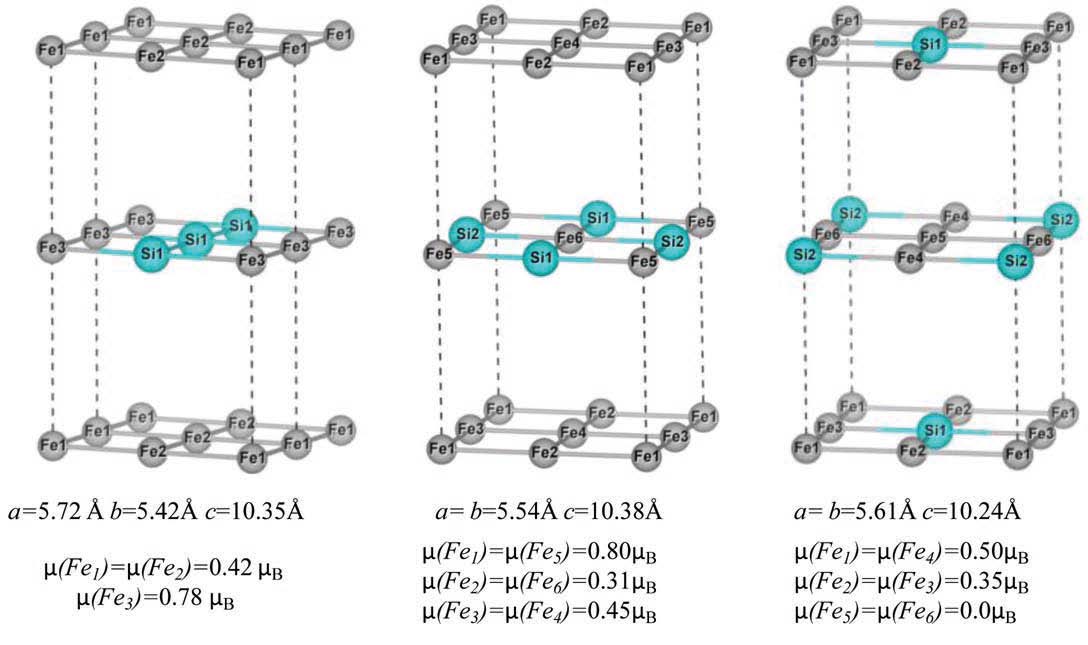}

\caption{(\emph{Color online}) Different environments for \emph{$Fe$} ions
leads to the formation of local MMs when a part of \emph{$Fe$} ions
are replaced by \emph{$Si$} ions (the host \emph{$Si$} atoms are
not shown). The optimized lattice parameters and the calculated MMs
of the \emph{Fe} atoms are given under the structures.\label{fig:Fig.10}}
\end{figure}

\end{widetext}

Fig. \ref{fig:Fig.10} displays three different variants of substitution
of \emph{Fe} atoms in the \emph{Fe} planes by the \emph{Si} atoms.
All calculations were carried out for the supercells $2a\times2b\times2c$
of \emph{$\alpha-FeSi_{2}$}, containing six iron atoms and two additional
\emph{$Si$} atoms. After full optimization of the supercells all
considered alloys become magnetic, but the magnitude of the magnetic
moment $\mu$ per supercell depends on the particular arrangement
of substitutional \emph{Si} atoms: $\mu=3.2\mu_{B}$ (Fig. \ref{fig:Fig.10}a),
$\mu=3.1\mu_{B}$ (Fig. \ref{fig:Fig.10}b), and $\mu=1.7\mu_{B}$
(Fig. \ref{fig:Fig.10}c). The emergence of local MM on different
\emph{$Fe$} atoms in the first two alloys (Fig. \ref{fig:Fig.10}a,b)
corresponds to the expectations, derived from the model. Indeed, since
the number of iron NNN surrounded $Fe_{3}$ atom \textcolor{black}{in
the first alloy} (Fig. \ref{fig:Fig.10}a) is decreased by two, the
local magnetic moment $\mu(Fe_{3})=0.8\mu_{B}$ on the $Fe_{3}$ atom
arises. Similar local MM appears on the $Fe_{1}$ and the $Fe_{5}$
atoms in the second alloy (Fig. \ref{fig:Fig.10}b) due to an increase
of the distance between NNN $Fe-Fe$ till $\backsimeq$2.8Å. The third
alloy (Fig. \ref{fig:Fig.10}c), however, presents an example where,
it seems, the model is oversimplified: the GGA calculation produces
zero moment on the $Fe_{5}$ atom without \emph{Fe} atoms in NNN surrounding,
while according to our model the biggest local magnetic moment have
to arise on the $Fe_{5}$ in this case. We assume that the term responsible
for it and which is missed in our model is the crystal electric field
(CEF), created by the \emph{Si} surrounding. The $Fe_{5}$ in the
third alloy (Fig. \ref{fig:Fig.10}c) sits in the most symmetrical
local surrounding $P4/mmm$ by \emph{Si} atoms, where the CEF splitting
has to be stronger than in the first two cases (Fig. \ref{fig:Fig.10}a,
b).

\textcolor{black}{The statement that the magnetic moments in $Fe-Si$
alloys can arise due to an increase of the $Si$ concentration allows
us to suggest an alternative explanation of the ferromagnetism in
the}\textcolor{black}{\emph{$\alpha-FeSi_{2}$}}\textcolor{black}{(111)
film on }\textcolor{black}{\emph{Si}}\textcolor{black}{(001) substrate,
successfully stabilized by the authors of \cite{key-16}. The authors
of Ref.\cite{key-16} explain the ferromagnetizm of the}\textcolor{black}{\emph{
$\alpha-FeSi_{2}$}}\textcolor{black}{{} film by the small concentration
(about 3\%) of the additional substitional Fe atoms. The calculation
in \cite{key-16}were performed in the framework of CPA, which is
not able to take into account the local-environment effects. We assume
that the observed moment arises not due to an increase of the }\textcolor{black}{\emph{Fe}}\textcolor{black}{{}
concentration as stated in the work \cite{key-16}, but due to an
increase of }\textcolor{black}{\emph{$Si$}}\textcolor{black}{{} concentration
that arises due to a diffusion of the}\textcolor{black}{\emph{ $Si$}}\textcolor{black}{{}
atoms from the }\textcolor{black}{\emph{$Si$ }}\textcolor{black}{substrate.
For example, the lattice parameters in the considered here }\textcolor{black}{\emph{$Si$}}\textcolor{black}{-rich
alloys are such that the (111) elementary-cells sizes of the}\textcolor{black}{\emph{
$\alpha-FeSi_{2}$}}\textcolor{black}{{} are very close to the $Si(001)-(3\times2)(11.5\times7.68\mathring{A})$:
$(11.83\times7.89\mathring{A}),\:(11.76\times7.83\mathring{A}),\:(11.67\times7.93\mathring{A})$
for the first (Fig. \ref{fig:Fig.10}a), second (Fig. \ref{fig:Fig.10}b)
and third (Fig.\ref{fig:Fig.10}c) alloys, correspondingly. This corresponds
to the mismutch about ( -1.5\% )-( -2.5\% ). Such a low mismutch presents
an opportunity to stabilize the epitaxial films of the }\textcolor{black}{\emph{$\alpha-FeSi_{2}$}}\textcolor{black}{{}
structure with similar arrangments of }\textcolor{black}{\emph{$Si$}}\textcolor{black}{{}
atoms. The magnetic moment $\mu\eqsim0.2-:-0.4\mu_{B}/f.u$. arises
for all types of the substitutions shown in Fig. \ref{fig:Fig.10},
which is consistent with the observed in Ref.\cite{key-16} values. }

\section{CONCLUSIONS}

\textcolor{red}{.}\textcolor{black}{Today it is recognized, that the
large, if not the decisive, role in the mechanism of the magnetic
structure formation in different compounds is played by the local
environment of the magnetic species. However, most of }\textcolor{black}{\emph{ab
initio}}\textcolor{black}{{} codes, based on DFT, are complicated and
usually represents a ``black box'', impedes the physical interpretation
of the results. In particularly, it is difficult to extract the contributions
from different local environment of an atom. Effects of local environment
are especially important in alloys, in which the slight difference
in the local environment can result in significantly different magnetic
structures. The substitutionally disordered systems such as metallic
alloys play an increasingly important role in technological applications
and, hence, a lot of efforts are invested into a theoretical understanding
of their properties. Although CPA is nowadays the most successful
}\textcolor{black}{\emph{ab initio }}\textcolor{black}{theory}\textcolor{black}{\emph{
}}\textcolor{black}{for the calculations of disordered alloys, the
standard formulation of it neglects the effects of NNN environment.
Along with the development of the }\textcolor{black}{\emph{ab initio}}\textcolor{black}{{}
methods (as non local CPA\cite{key-33}), the understanding of the
certain property of the specific compound can be reached in the framework
of the suitable models with the parameters obtained from the }\textcolor{black}{\emph{ab
initio }}\textcolor{black}{calculations for a given compound. Namely
such combination of the }\textcolor{black}{\emph{ab initio}}\textcolor{black}{{}
calculations with the multiorbital model one for the iron silicides
was used in our work. The feature which distiguish our model approach
from other ones is that the parameters of the model are determined
from the fitting its self-consistent charge density to the one, obtained
by }\textcolor{black}{\emph{ab initio}}\textcolor{black}{{} calculations.
This allows to study the effects of NN and NNN local environment of
the }\textcolor{black}{\emph{$Fe$}}\textcolor{black}{{} atoms on the
MM formation. }

\textcolor{black}{The presented study of the effect of silicon-atoms'
substitution by the iron atoms and }\textcolor{black}{\emph{vice versa
}}\textcolor{black}{on the magnetic properties in the iron silicide
$\alpha-FeSi_{2}$ within the suggested multiorbital model has shown
that while the stoichiometric material $\alpha-FeSi_{2}$ is nonmagnetic,
the appearance of substitutional iron atoms in the $\alpha-FeSi_{2}$
may result in different magnetic structures, either ferromagnatic
or ferrimagnetic. Which particular structure emerges is determined
by the number and the }\textcolor{black}{\emph{spatial arrangement}}\textcolor{black}{{}
of the substitutional iron atoms. The latter statement is strongly
supported by the fact that different magnetic structures can appear
at }\textcolor{black}{\emph{the same}}\textcolor{black}{{} concentration
of substitutional }\textcolor{black}{\emph{Fe}}\textcolor{black}{{}
atoms. Besides, as follows from the Hartree-Fock model calculations,
the MMs formation is essentially determined not by the NN }\textcolor{black}{\emph{Si}}\textcolor{black}{{}
atoms but by the NNN environment, particularly, by the }\textcolor{black}{\emph{$Fe$}}\textcolor{black}{{}
atoms along the crystallgraphic axes: the MMs on iron atoms are very
sensitive to the values of NNN }\textcolor{black}{\emph{Fe - Fe}}\textcolor{black}{{}
hopping parameters $t_{3}$. We demonstrated it by a comparison of
the maps of moments dependence on the hopping parameters with and
without taking into account NNN ones. It is important that the nonmagnetic
states in the stoichiometric $\alpha-FeSi_{2}$ arise at NNN $t_{3}\neq0$
only. The model with NN hoppings only, even if all NN to }\textcolor{black}{\emph{Fe}}\textcolor{black}{{}
atoms are }\textcolor{black}{\emph{Si}}\textcolor{black}{{} atoms, does
not have the solutions with zero moments on }\textcolor{black}{\emph{Fe}}\textcolor{black}{.
This allows to suggest that the magnetism in the nonmagnetic $\alpha-FeSi_{2}$
can be induced by a negative pressure. }

\textcolor{black}{The various magnetic structures (ferro-, ferri-
or nonmagnetic) in }\textcolor{black}{\emph{Fe}}\textcolor{black}{-rich
alloys also controlled by the NNN }\textcolor{black}{\emph{Fe - Fe}}\textcolor{black}{{}
hopping parameters $t_{3}$. The different ways of }\textcolor{black}{\emph{Si}}\textcolor{black}{{}
atoms substitution by }\textcolor{black}{\emph{Fe}}\textcolor{black}{{}
atoms result in the diverse local distortions of the underlying lattice
and, in turn, to quite different hopping parameters $t_{3}$ and magnetic
properties. It is most clearly demostrated by the magnetic behavior
of several alloys with the same concentration of substitutional atoms,
}\textcolor{black}{\emph{e.g.}}\textcolor{black}{, alloys }\textbf{\textcolor{black}{C}}\textcolor{black}{{}
and }\textbf{\textcolor{black}{D}}\textcolor{black}{{} considered in
this work (Sec.IIB). The comparison of the magnetic-moments maps reveals
that different values of NNN }\textcolor{black}{\emph{$Fe-Fe$}}\textcolor{black}{{}
hopping parameters $t_{3}$ lead to the diverse magnetic behaivour:
a nonmagnetic one in the alloy }\textbf{\textcolor{black}{C}}\textcolor{black}{{}
and a ferromagnetic one in the alloy }\textbf{\textcolor{black}{D}}\textcolor{black}{.
Notice, despite of the different lattice distortion, the spacing between
NN (as well as number of NN }\textcolor{black}{\emph{Si }}\textcolor{black}{atoms)
is the same in the both cases, hence, the local environment models
which do not take into account the NNN hoppings, cannot explain this
distinction. Unlike the local environment models \cite{key-18,key-19}
we observe a weak dependence of the }\textcolor{black}{\emph{Fe}}\textcolor{black}{{}
magnetic moment on the hopping $t_{2}$ between NN }\textcolor{black}{\emph{$Fe$}}\textcolor{black}{{}
and }\textcolor{black}{\emph{$Si$}}\textcolor{black}{{} atoms: all
the $t_{1}-t_{2}$-maps for }\textcolor{black}{\emph{$Fe$}}\textcolor{black}{{}
moments, calculated within this model, are elongated along the axis
$t_{2}$. One more characteristic feature of these maps is the presence
of sharp boundaries between magnetic states region and nonmagnetic
one as a function of NN }\textcolor{black}{\emph{$Fe-Fe$}}\textcolor{black}{{}
hopping integral $t_{1}$. In general, our conclusion about the decisive
role of NNN local environment in the magnetic moment formation contradicts
to the conclusions of earlier (much less detailed) models of local
environment, where a decrease of the moment on }\textcolor{black}{\emph{$Fe$}}\textcolor{black}{{}
atoms was ascribed to the increase of number of }\textcolor{black}{\emph{$Si$}}\textcolor{black}{{}
in NN sphere. According to our calculations the main role in formation
of local magnetic moment is played by decreasing of the number of
}\textcolor{black}{\emph{$Fe-Fe$}}\textcolor{black}{{} pairs along
the crystallographic axes and/or increasing of the distance between
them. This conclusion is especially interesting since most of models
do not take the NNN hoppings into account. }

\textcolor{black}{The unexpected and somewhat counter-intuitive conclusion,
produced by the model calculations, is that not only an increase of
the }\textcolor{black}{\emph{Fe}}\textcolor{black}{{} concentration
can lead to the emergence of local magnetic moment on }\textcolor{black}{\emph{Fe}}\textcolor{black}{{}
atoms, but also of the metalloid concentration. Indeed, the number
of the}\textcolor{black}{\emph{ $Fe-Fe$ }}\textcolor{black}{pairs
can be reduced by replacing of the }\textcolor{black}{\emph{Fe }}\textcolor{black}{atoms
in iron planes by }\textcolor{black}{\emph{$Si$}}\textcolor{black}{{}
atoms. Moreover the distances between}\textcolor{black}{\emph{ Fe
}}\textcolor{black}{atoms in these planes are increased due to the
distortion of the underlying lattice. So, the conditions leading to
the emergence of magnetism are met. The }\textcolor{black}{\emph{ab
initio}}\textcolor{black}{{} calculation of the ordered $Si$-rich alloys
confirms this conclusion. Hence we can explain the ferromagnetism
in the $\alpha-FeSi_{2}(111)$ film, obtained by the authors of Ref.\cite{key-16},
in a more realistic way. In our opinion, the observed in Ref.\cite{key-16}
moment results from the increase of }\textcolor{black}{\emph{$Si$}}\textcolor{black}{{}
concentration due to a diffusion of the }\textcolor{black}{\emph{$Si$}}\textcolor{black}{{}
atoms from the }\textcolor{black}{\emph{$Si$ }}\textcolor{black}{substrate,
but not due to an increase of the }\textcolor{black}{\emph{$Fe$}}\textcolor{black}{{}
concentration}\textcolor{red}{. }

\textcolor{black}{Based on the presented analysis, we can formulate
the conditions promoting the appearence of a magnetism in the iron
silicides. The key parameters responsible for the magnetism are the
hoppings between }\textcolor{black}{\emph{$Fe$}}\textcolor{black}{{}
atoms $t_{1}$ and $t_{3}$, which are the most sensitive parameters
to different types of pressure. The latter can be done by either by
fitting the lattice parameter of the substrate for $\alpha-FeSi_{2}$
film (chemical pressure), or by a sustitution of }\textcolor{black}{\emph{$Fe$}}\textcolor{black}{{}
or }\textcolor{black}{\emph{$Si$}}\textcolor{black}{{} atoms. As was
pointed out in Ref.\cite{key-17}, the best orientation relationships,
that stabilize the epitaxial $\alpha-FeSi_{2}$ are $\alpha-FeSi_{2}(201)||Si(110)$,
$\alpha-FeSi_{2}(110)||Si(110)$ or $\alpha-FeSi_{2}(111)||Si(001)$.
Such planes contain additional }\textcolor{black}{\emph{$Si$}}\textcolor{black}{{}
atoms in }\textcolor{black}{\emph{$Si$}}\textcolor{black}{-rich alloys
from Fig. \ref{fig:Fig.10} and the sizes of corresponding unit cells
are very close to the }\textcolor{black}{\emph{$Si$}}\textcolor{black}{-substrate
one. Small mismatch has place for the all mutual orientations of film
and substrate and presents an opportunity to stabilize the epitaxial
films of the $\alpha-FeSi_{2}$ structure. Moreover the possibility
of tuning the hopping parameter $t_{3}$ in iron silicides has the
large tecnological interest, because it gives an opportutity to control
the appeance of different magnetic configurations in the cause of
fabrication of new alloys or nanostructures with the prospective magnetic
properties. At last, the existence of the region with sharp transition
from ferro- to paramagnetic or from ferro- to ferrimagnetic state
strongly improves the perspectives of the practical applications of
iron silicide films and, hopefully, will stimulate technologists to
find a way to make the films near the instability line with desirable
characteristics.}
\begin{acknowledgments}
This work was supported by the Russian Foundation for Basic Research,
projects No 14-02-00186, 17-02-00161 and by the joint Krasnoyarsk
regional scientific foundation and Russian Foundation for Basic Research,
projects No 16-42-242036, 16-42-243035. The authors would like to
thank AS Shinkorenko for the technical support.\end{acknowledgments}

\end{document}